\documentclass[12pt]{article}

\usepackage{hyperref}

\usepackage{graphicx}
\usepackage{appendix}
\usepackage{amsmath, amssymb, mathrsfs}
\usepackage{tikz, tcolorbox, color}
\usetikzlibrary{decorations.markings, arrows.meta, calc}
\tcbuselibrary{skins}
\usepackage {diagbox}
\usepackage{ulem}
\usepackage{physics}
\usepackage{fancybox}
\usepackage{fancyhdr}
\usepackage{bm}

\usepackage{epsf}
\usepackage{amsmath,amssymb}
\usepackage{latexsym}
\usepackage{graphicx,color}
\usepackage{wrapfig}
\usepackage{latexsym}
\usepackage{graphics}

\usepackage{color}
\usepackage{delarray,color}
\usepackage{fancybox}  

\usepackage[T1]{fontenc}
\usepackage{lmodern}
\usepackage[utf8]{inputenc}

\usepackage{amsmath,amssymb,mathtools,bm}
\usepackage{physics}
\usepackage{tikz}
\usepackage{hyperref}

\newcommand {\beq}{\begin{align}}
\newcommand {\eeq}{\end{align}}

\newcommand{\be}{\begin{equation}}
\newcommand{\ba}{\begin{align}}
\newcommand{\ea}{\end{align}}
\newcommand{\ee}{\end{equation}}

\newcommand{\s}{\sqrt}

\newcommand{\beqa}{\begin{align}}
\newcommand{\eeqa}{\end{align}}

\newcommand{\unit}{\hbox to 3.8pt{\hskip1.3pt \vrule height 7.4pt
    width .4pt \hskip.7pt \vrule height 7.85pt width .4pt \kern-2.4pt
    \hrulefill \kern-3pt \raise 3.7pt\hbox{\char'40}}}

\def\matt[#1,#2,#3,#4]{\left(%
\begin{array}{cc} #1 & #2 \\ #3 & #4 \end{array} \right)}

\usepackage{tabularx}

\catcode`\@=11
\@addtoreset{equation}{section}

\catcode`@=12
\relax

\begin{document}
%%%%%%%%%%%%%%%%%%%%%%%%%%%%%%%%%%%%%%%%%%%%%%%%%%%%%%%%%%%%%%%%%%%%%%%%
%\baselineskip 0.7cm

\begin{titlepage}

%% Set the number of the title with 0
\setcounter{page}{0}

%% change the footnote symbol
\renewcommand{\thefootnote}{\fnsymbol{footnote}}

\begin{flushright}
%{\tt 
YITP-26-34

KUNS-3097
%\\}
\end{flushright}

\vskip 1.35cm

\begin{center}
{\Large \bf 
Rindler Physics with a UV Cutoff on the Lattice
}

\vskip 1.2cm 

{\normalsize
Seiken  Chikazawa${}^{a}$~\footnote{chikazawa(at)gauge.scphys.kyoto-u.ac.jp} and Seiji Terashima${}^{b}$~\footnote{terasima(at)yukawa.kyoto-u.ac.jp}
}

\vskip 0.8cm

${}^{a}${\it
Department of Physics, Kyoto University,
Kyoto 606-8502, Japan}

${}^{b}${\it
Center for Gravitational Physics and Quantum Information, \\
Yukawa Institute for Theoretical Physics, Kyoto University, Kyoto 606-8502, Japan
}

\end{center}

\vspace{12mm}

\centerline{{\bf Abstract}}

We investigate quantum field theory in Rindler space with a UV cutoff by considering a free scalar field on a lattice in Rindler coordinates. %On the lattice, the local Rindler Hamiltonian does not coincide with the modular Hamiltonian of the Minkowski vacuum, which becomes nonlocal once the cutoff is introduced. We study the consequences of this distinction.
We find that the Minkowski vacuum is not exactly thermal with respect to the local lattice Rindler Hamiltonian. Nevertheless, for observables sufficiently far from the horizon, the Wightman function and the Unruh--DeWitt detector response reproduce the expected thermal behavior in the continuum limit. Thus, the Unruh effect survives operationally, even though exact thermality is lost at the state level. We also show that the Rindler vacuum energy density reproduces the standard continuum behavior away from the horizon, while the UV singularity at the horizon is replaced by a stretched-horizon contribution.
Furthermore, the retarded Green function exhibits a component reflected at the stretched horizon, implying that an ingoing wave packet is reflected at a proper distance of order the cutoff. This provides an effective brick-wall picture in the UV-regulated theory. Our analysis suggests that, once a cutoff is introduced, the global Minkowski description and the wedge description based on a local Rindler Hamiltonian are no longer equivalent at the operator level.

\end{titlepage}

\newpage

\tableofcontents
\vskip 1.2cm

\section{Introduction and Summary}

A useful starting point for understanding black hole horizon physics is the Rindler description of Minkowski spacetime \cite{Rindler:1966zz, Unruh:1976db, Hawking:1975vcx}. In the near-horizon region of a (nonextremal) black hole, if one focuses only on scales much shorter than the curvature scale, the exterior geometry is locally approximated by a Rindler wedge. The right wedge describes uniformly accelerated observers in \(x>0\), and together with the left wedge, related by parity, covers the whole Minkowski spacetime. In the continuum theory, the Minkowski vacuum takes the form of a thermofield-double state with respect to the left and right Rindler wedges, and tracing out the left wedge yields a thermal density matrix on the right wedge \cite{Unruh:1976db, Israel:1976ur, Bisognano:1976za}.

In this paper, we reconsider this familiar picture in a quantum field theory with a UV cutoff. Physically, one does not know the theory at arbitrarily high energy scales, and one should therefore ask which aspects of the usual continuum near-horizon physics are robust under UV regularization. More precisely, if low-energy observables were strongly sensitive to the details of trans-cutoff physics, then their predictive power would be lost. It is therefore important to understand whether the standard continuum results in Rindler space surviving after a UV cutoff is introduced. Since the Rindler coordinate system is singular at the horizon, this question is not entirely trivial.

We study this issue concretely in the \((1+1)\)-dimensional massless free scalar field, with the UV cutoff implemented by lattice regularization in the Hamiltonian formalism on Minkowski spacetime.\footnote{For the spin system (or fermion) model of Rindler physics, see \cite{Okunishi:2019dmv, Kinoshita:2024ahu, Yang:2019kbb, Shi:2021nkx}, which mainly considered the entanglement entropy.} From the viewpoint of the Rindler wedge, this regularization effectively introduces a stretched-horizon or brick-wall-like structure. In particular, there is no degree of freedom exactly at the horizon, and the nearest lattice site lies at a finite proper distance from it. In this sense, the UV regularization by the lattice introduces an effective brick wall: degrees of freedom inside the stretched horizon are absent, and the horizon singularity of the continuum theory is replaced by a finite near-horizon structure. Even when one considers the full left-plus-right system, we are interested in the theory restricted to the right wedge, obtained by tracing out the left wedge.

The purpose of introducing such a UV-regulated model is to identify which parts of the idealized continuum Rindler physics are UV-sensitive. In the continuum theory, the Rindler Hamiltonian coincides with the Lorentz boost generator and also with the modular Hamiltonian of the Minkowski vacuum restricted to a wedge. On the lattice, however, exact Lorentz symmetry is lost, and these notions no longer coincide. In particular, the exact modular Hamiltonian becomes nonlocal. In this paper, we instead adopt the natural local lattice Hamiltonian in Rindler coordinates. 
%(with a Neumann-type boundary condition at the endpoint).\footnote{We expect that the results in this paper do not depend on the boundary condition we choose.}
This choice preserves manifest lattice ultralocality and reproduces, in the continuum limit %large-\(N\) limit 
and away from the horizon, the expected local Rindler Hamiltonian.\footnote{We also assume that the Rindler Hamiltonian is free.}

Within this lattice model, we show numerically that for the lattice version of the Minkowski vacuum, the Wightman function with respect to the Rindler Hamiltonian agrees, far from the horizon and in the continuum limit, with the finite-temperature Wightman function expected from the Unruh effect \cite{Unruh:1976db}. Equivalently, an Unruh-DeWitt detector coupled to such observables responds thermally. In this operational sense, the Unruh effect is insensitive to the UV cutoff. At the same time, however, the Minkowski vacuum is not exactly a thermal state with respect to the lattice Rindler Hamiltonian. This difference is substantial and UV-dependent. The reason is also physically natural: the continuum Rindler modes are dominated by high-energy Minkowski components \cite{Sugishita:2022ldv}, and their exact thermal structure is therefore sensitive to the UV completion. Thus, in the cutoff theory, low-energy or local probes can still exhibit thermal behavior, while exact thermality as a state with respect to the Rindler Hamiltonian is lost.

We also compute the energy density of the Rindler vacuum \cite{Fulling:1972md, Boulware:1974dm} in this UV-regulated model. Away from the horizon, we reproduce the well-known continuum behavior proportional to \(-1/x^2\)~\cite{Candelas:1977zza}. Near \(x=0\), however, the lattice regularizes the singularity into an \(O(1/a^2)\) contribution, where \(a\) is the lattice spacing, and the total energy is correspondingly \(O(1/a^2)\). This is consistent with the continuum analysis of the Rindler vacuum~\cite{Parentani:1993yz}, where energy density is positive and %proportional to $\delta(0)$ 
divergent at $x=0$, 
while making explicit that the singular horizon contribution is spread over the stretched-horizon region once the UV cutoff is introduced.

Furthermore, we show that the retarded Green function for the lattice Rindler Hamiltonian has a contribution reflected at the brick wall. Since the retarded Green function in the free theory is state-independent, this is a direct dynamical manifestation of the near-horizon cutoff structure. Correspondingly, a wave packet propagating toward the horizon is strongly reflected in the region \(x=O(a)\), which we call the brick wall. These results are consistent with the expectation that UV-cutoff effects are invisible to probes localized away from the horizon at short Rindler times. However, because signals reflected from the near-horizon region return to their original location after a Rindler time of order
\begin{equation}
t_R \sim \frac{\ln (x/a)}{\kappa} \sim 2 \pi \frac{\ln (x/ a)}{T_H},
\label{st}
\end{equation}
with \(\kappa\) the Rindler acceleration parameter and 
$T_H$ is the temperature, 
%\(N\) the number of lattice sites, 
the effect of the UV cutoff can eventually become visible even to observables located far from the horizon.

The broader motivation for this analysis comes from black-hole physics. One would like to understand whether there is, in principle, any observable difference between an eternal black hole and a black hole formed by collapse after sufficiently long time evolution, even at the level of low-energy correlation functions. It is often implicitly assumed that, for such low-energy observables, these two situations behave essentially in the same way (see \cite{KawamotoST} for some discussions against such an assumption). In AdS/CFT \cite{Maldacena:1997re}, this question is often phrased in terms of the difference between the canonical ensemble on one CFT, obtained by tracing out one side of the thermofield-double state on \( \mathrm{CFT}_1 \times \mathrm{CFT}_2 \) \cite{Maldacena:2001kr}, and a typical pure state in a single CFT. The corresponding Rindler analogue is the contrast between the Minkowski vacuum (or the associated thermal state on the right wedge) and generic pure states on the right wedge.

The significance of the present lattice model for black-hole physics is the following. At finite \(N\) in AdS/CFT, the bulk theory cannot literally contain the arbitrarily short-distance degrees of freedom of continuum effective field theory. Since the near-horizon region is precisely where UV structure is most strongly probed because of the large blueshift, one expects some effective loss of continuum degrees of freedom there. This should manifest itself as a brick-wall-like \cite{tHooft:1984kcu} or stretched-horizon \cite{Susskind:1993if} structure.\footnote{The possibility of the breakdown of the equivalence principle near the horizon was argued in \cite{Mathur:2005zp} \cite{Almheiri:2012rt} from the non-singular gravity solutions and the discussion of quantum information.} Our lattice model is not proposed as the microscopic bulk theory itself; rather, it is a concrete toy model of a horizon region in which the number of effective UV degrees of freedom is finite. In this sense, the lattice realization captures an aspect that should be generic whenever finite-\(N\) quantum gravity removes the continuum near-horizon mode structure.

There have also been more direct discussions in the literature suggesting brick-wall-like structures in finite-\(N\) AdS/CFT~\cite{Iizuka:2013kma}, as well as arguments from bulk reconstruction that bulk degrees of freedom inside such a stretched horizon should not exist as independent semiclassical degrees of freedom~\cite{Terashima:2021klf, Sugishita:2022ldv, Sugishita:2023wjm, Sugishita:2024lee, Terashima:2025shl} based on the operator formalism of AdS/CFT \cite{Terashima:2017gmc, Terashima:2019wed, Terashima:2020uqu}. 
The present analysis provides a simple and explicit realization of how such a finite-\(N\) cutoff structure can affect real-time observables. In particular, if one assumes that finite-\(N\) black holes effectively possess such a near-horizon brick-wall-like structure, then our results suggest that low-energy correlators may begin to deviate from the eternal-black-hole or exact thermal answer after times of order \(t_R \sim \ln (x/a)/\kappa\), because of reflected contributions from the stretched horizon.

This possibility is potentially important. The usual continuum argument that the Minkowski energy density diverges positively at \(x=0\) for the Rindler vacuum \cite{Parentani:1993yz}, and more generally for Rindler energy eigenstates~\cite{Czech:2012be}, might appear to suggest a large distinction between thermofield-double-like states and generic right-wedge states. However, in the strict continuum theory that singularity is localized at \(x=0\) itself and behaves essentially as a horizon-supported positive divergence. From the viewpoint of observables entirely inside the right wedge, it does not by itself imply a difference in ordinary low-energy bulk physics away from the horizon. Once the UV structure is made finite, however, the singular contribution is spread over the stretched-horizon region, and reflected signals can re-enter the observable region after a parametrically long but finite Rindler time. This is the mechanism by which finite UV resolution, which is a finite-$N$ effect and a finite Planck scale effect in AdS/CFT, can make a physically meaningful difference.\footnote{This difference will disappear if we take $N=\infty$ formally or $1/N$ expansion because the r.h.s. of \eqref{st} diverges for $a \rightarrow 0$ and $a$ can be regarded as the Planck length.}

If such a mechanism persists in the presence of gravitational interactions, as has been argued in \cite{Terashima:2025tct} in discussions of black-hole echoes in the gravitational wave observations~\cite{Cardoso:2019rvt, Abedi:2020ujo}, then finite-\(N\) effects of this kind may be relevant not only conceptually but also observationally. In this sense, the lattice Rindler model provides a simple setting in which one can separate those aspects of horizon thermality that are robust under UV regularization from those that are not, and at the same time obtain a concrete toy model for possible finite-\(N\) horizon effects in black-hole physics.

\section{Lattice field theory in Rindler Spacetime}

In this section, we discretize two-dimensional Rindler spacetime into a lattice and consider the quantum field theory of a massless scalar field on it.  
In subsection~\ref{2.1}, we quantize the scalar field, and in subsection~\ref{2.2}, we compute the vacuum expectation value of the energy density $T_{00}$ of the scalar field.  
In subsection~\ref{2.3}, we calculate the derivative of the retarded Green's function and show that a stretched horizon effectively emerges.  
In subsection~\ref{2.4}, we evaluate the expectation value of the particle number produced by the Unruh effect in order to see how the thermal distribution obtained in the continuum theory is modified.

\subsection{Quantization %in Lattice Rindler Spacetime
}\label{2.1}

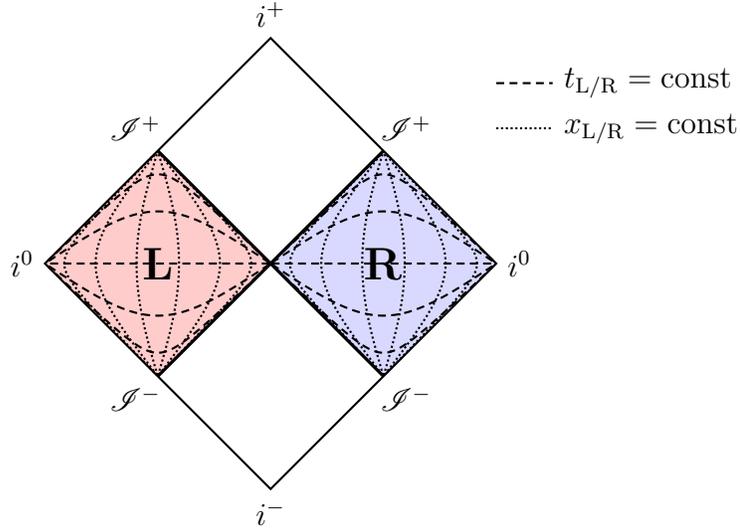
\begin{figure}[h]
\centering
\begin{tikzpicture}[scale=1.5]
    \fill[blue, fill opacity=0.15] (0,0) -- (1,1) -- (2,0) -- (1,-1) -- cycle;
    \fill[red, fill opacity=0.20] (0,0) -- (-1,1) -- (-2,0) -- (-1,-1) -- cycle;

    \draw[thick] (0,-2) node[below] {$i^-$} 
              -- (2,0)  node[right] {$i^0$} 
              -- (0,2)  node[above] {$i^+$} 
              -- (-2,0) node[left]  {$i^0$} 
              -- cycle;
              
    \node at (1.2, 1.2) {$\mathscr{I}^+$};
    \node at (1.2, -1.2) {$\mathscr{I}^-$};
    \node at (-1.2, 1.2) {$\mathscr{I}^+$};
    \node at (-1.2, -1.2) {$\mathscr{I}^-$};

    \draw[very thick] (-1,-1) -- (1,1);
    \draw[very thick] (-1,1) -- (1,-1);

    \node at (1, 0) {\Large \textbf{R}};
    \node at (-1, 0) {\Large \textbf{L}};

    \foreach \s in {-2, -1.0, -0.3, 0.3, 1.0, 2} {
        \draw[thick, densely dotted] plot [smooth, domain=-5:5, samples=100] 
            ({ (atan(exp(\s+\x)) + atan(exp(\s-\x)))/90 }, 
             { (atan(exp(\s+\x)) - atan(exp(\s-\x)))/90 });
    }

    \foreach \t in {-1.8, -0.8, 0, 0.8, 1.8} {
        \draw[thick, densely dashed] plot [smooth, domain=-5:5, samples=100] 
            ({ (atan(exp(\x+\t)) + atan(exp(\x-\t)))/90 }, 
             { (atan(exp(\x+\t)) - atan(exp(\x-\t)))/90 });
    }

    \foreach \s in {-2, -1.0, -0.3, 0.3, 1.0, 2} {
        \draw[thick, densely dotted] plot [smooth, domain=-5:5, samples=100] 
            ({ -(atan(exp(\s+\x)) + atan(exp(\s-\x)))/90 }, 
             { (atan(exp(\s+\x)) - atan(exp(\s-\x)))/90 });
    }

    \foreach \t in {-1.8, -0.8, 0, 0.8, 1.8} {
        \draw[thick,  densely dashed] plot [smooth, domain=-5:5, samples=100] 
            ({ -(atan(exp(\x+\t)) + atan(exp(\x-\t)))/90 }, 
             { (atan(exp(\x+\t)) - atan(exp(\x-\t)))/90 });
    }

    \begin{scope}[shift={(2, 1.6)}]
        \draw[thick, densely dashed] (0,0) -- (0.5,0) node[right, text=black] {$t_{\mathrm{L/R}} = \text{const}$};
        \draw[thick, densely dotted] (0,-0.4) -- (0.5,-0.4) node[right, text=black] {$x_{\mathrm{L/R}} = \text{const}$};
    \end{scope}

\end{tikzpicture}
\label{Penrose}
\caption{Penrose diagram of Rindler spacetime.}
\end{figure}

In Minkowski spacetime $\dd{s^2}=-\dd{T}^2+\dd{X}^2$, two Rindler wedges, the left wedge and the right wedge, can be defined.  
Their coordinates $\{t_{\mathrm{L}},x_\mathrm{L}\}$ and $\{t_{\mathrm{R}},x_\mathrm{R}\}$ are given, using a constant $\kappa$, by
\begin{align}
\begin{cases}
T=-x_{\mathrm{L}}\sinh\kappa t_{\mathrm{L}} \\
X=-x_{\mathrm{L}}\cosh\kappa t_{\mathrm{L}}
\label{coordinateL}
\end{cases}
\end{align}
and
\begin{align}
\begin{cases}
T=x_{\mathrm{R}}\sinh\kappa t_{\mathrm{R}} \\
X=x_{\mathrm{R}}\cosh\kappa t_{\mathrm{R}}
\label{coordinateR}
\end{cases}
\end{align}
respectively.  
Here $ %t_\mathrm{L/R}>0,\,\,\,
x_\mathrm{L/R}>0$.  
The metric of Rindler spacetime is
\begin{align}
\dd{s^2}=-\kappa^2 x_\mathrm{L/R}^2\dd{t_\mathrm{L/R}^2}+\dd{x_\mathrm{L/R}^2},
\end{align}
which has a horizon at $x_\mathrm{L/R}=0$.  
The Penrose diagram is shown in Fig.~\ref{Penrose}.  
In what follows, unless otherwise stated, we focus on the right wedge.  
For simplicity, we omit the subscript R and denote the coordinates simply by $\{t, x\}$.

Let us consider a massless free scalar field $\phi$ on Rindler spacetime.  
The action is
\begin{align}
S&=-\frac{1}{2}\int\dd^2{x}\sqrt{-g}g^{\mu\nu}\partial_\mu\phi\,\partial_\nu\phi \\
&=\int\dd^2{x}\left[\frac{1}{2\kappa x}(\partial_t\phi)^2-\frac{\kappa x}{2}(\partial_x\phi)^2\right]\equiv \int\dd^2{x}\,\mathcal{L}, 
\end{align}
and the canonical momentum $\pi$ conjugate to $\phi$, the Hamiltonian density $\mathcal{H}$, and the Hamiltonian $H$ are
\begin{align}
&\pi=\pdv{\mathcal{L}}{(\partial_t\phi)}=\frac{1}{\kappa x}\partial_t\phi,\\
&\mathcal{H}=\pi(\partial_t\phi)-\mathcal{L}=\frac{1}{2}\kappa x\,\pi^2+\frac{1}{2}\kappa x\,(\partial_x\phi)^2, \\
&H=\int\dd{x}\,\mathcal{H}=\int\dd{x}\,\left[\frac{1}{2}\kappa x\,\pi^2+\frac{1}{2}\kappa x\,(\partial_x\phi)^2\right]. 
\end{align}

We now discretize the $X$ direction as $X\to X_n=\left(n-\frac{1}{2}\right)a\,\,\,(n\in\mathbb{Z}_{>0})$ in order to obtain the usual lattice field theory in the Minkowski spacetime. Here $a$ is the lattice spacing. 
The lattice points are shifted by $1/2$ from the integers so that no point lies exactly on the horizon in the Rindler coordinate at $t_R=0$. We will later assume $n=1,2, \cdots, N$, which corresponds to 
introducing an IR cut-off.\footnote{We introduce $N \gg 1$ as an infrared (IR) regulator. We assume that the parameter $\kappa$, which is proportional to the temperature, satisfies $\kappa a \ll 1$. Furthermore, the energy scale under consideration can be regarded as being on the order of $\kappa$ (rather than $1/a$).}

For the corresponding lattice field theory in the (right) Rindler wedge,
we need to discretize the $x$ direction as $x\to x_n=\left(n-\frac{1}{2}\right)a\,\,\,(n\in\mathbb{Z}_{>0})$ at $t_R=0$.   
After discretization, the integral becomes $\int\dd{x}\to a\sum$.  
Writing $\phi_n=\phi(x_n)$ and $\pi_n=\pi(x_n)$, the discretized Hamiltonian becomes
\begin{align}
H=\sum_{n=1}^N\left[\frac{1}{2}\kappa a^2 \left(n-\frac{1}{2}\right)\pi_n^2+\frac{1}{2}\kappa a^2 n\frac{(\phi_{n+1}-\phi_{n})^2}{a^2}\right].
\label{RHamiltonian}
\end{align}
When discretizing the coefficient $\kappa x$ appearing in $\kappa x\,(\partial_x\phi)^2$, one must be careful about where it should be evaluated.  
Since the derivative $\partial_x\phi$ is replaced by the finite difference $(\phi_{n+1}-\phi_n)/a$, it is natural to evaluate $\kappa x$ at the midpoint between $x_{n+1}$ and $x_n$, namely $x_{n+\frac{1}{2}}=an$.  
Indeed, if the coefficient is not evaluated at the midpoint, the discrete version of integration by parts,
\begin{align}
\sum_{n}\kappa x_{n+\frac{1}{2}}\frac{(\phi_{n+1}-\phi_n)^2}{a^2}=-\sum_{n}\phi_{n}\frac{1}{a}\left(\kappa x_{n+\frac{1}{2}}\frac{\phi_{n+1}-\phi_n}{a}-\kappa x_{n-\frac{1}{2}}\frac{\phi_{n}-\phi_{n-1}}{a}\right)
\end{align}
does not hold, and results consistent with the continuum theory cannot be obtained.

Imposing the canonical commutation relation\footnote{In the continuum theory the delta function satisfies $\int\dd{x}\delta(x)=1$. The discretized version of this relation is $a\sum_{n}\delta_{m,n}/a=1$, so the quantity corresponding to $\delta(x)$ in the continuum theory is $\delta_{m,n}/a$.}
\begin{align}
\left[\phi_m, \pi_n\right]=i\delta_{m,n}/a, 
\end{align}
we quantize the system and determine the eigenstates of the Hamiltonian \eqref{RHamiltonian}.
Let us define the dimensionless variables:
\begin{align}
\tilde{\phi}_n=\frac{1}{\sqrt{n-\frac{1}{2}}}\phi_n, \qquad\tilde{\pi}_n=a \sqrt{n-\frac{1}{2}}\,\pi_n . 
\end{align}
Then \eqref{RHamiltonian}, with the IR cut-off, can be written as
\begin{align}
H=\frac{1}{2}\kappa %a^2
\sum_{n=1}^N\left[\tilde{\pi}_n^2+% \frac{1}{a^2}
\sum_{m=1}^N\tilde{\phi}_mA_{m,n}\tilde{\phi}_n\right]. 
\label{RHamiltonian2}
\end{align}
Here $A_{m,n}$ is the symmetric matrix
\begin{align}
A_{m,n}=\sqrt{\left(m-\frac{1}{2}\right)\left(n-\frac{1}{2}\right)}\left[(2n-1)\delta_{m,n}-n\delta_{m,n+1}-m\delta_{m+1,n}\right].
\end{align}
The field $\phi_{N+1}$, which appears due to the finite difference, is removed by imposing the boundary condition $\phi_{N+1}=0$.
Let $P$ be an orthogonal matrix satisfying $PAP^T=\mathrm{diag}\,(\lambda_n^2)$.  
Defining
\begin{align}
\Pi_n=\sum_{m=1}^N P_{n,m}\tilde{\pi}_m, \qquad\Phi_n=\sum_{m=1}^N P_{n,m}\tilde{\phi}_m, 
\end{align}
the Hamiltonian \eqref{RHamiltonian2} becomes
\begin{align}
H=\frac{1}{2}\kappa %a^2
\sum_{n=1}^N \left(\Pi_n^2+\lambda_n^2 \Phi_n^2\right). 
\end{align}
This is the Hamiltonian of harmonic oscillators. Introducing the creation and annihilation operators
\begin{align}
a_n^\dag=\frac{1}{\sqrt{2\lambda_n}}\left(\Pi_n+i \lambda_n \Phi_n\right),\qquad a_n=\frac{1}{\sqrt{2\lambda_n}}\left(\Pi_n-i \lambda_n \Phi_n\right), 
\end{align}
the Hamiltonian can be written as
\begin{align}
H=\sum_{n=1}^N \kappa\lambda_n\left(a_n^\dag a_n+\frac{1}{2}\right).
\end{align}
It is straightforward to verify that these operators satisfy the commutation relation
\begin{align}
[a_m, a_n^\dag]=\delta_{m,n}.
\end{align}
The (Rindler) vacuum state $\ket{0}_R$ of the scalar field $\phi$ in Rindler spacetime is defined by 
$a_n\ket{0}_R=0\,\,\, %(^\forall n)
$ 
for all the annihilation operators, and general excited states are given by $(a^\dag_1)^{n_1} (a^\dag_2)^{n_2} \cdots (a^\dag_N)^{n_N} \ket{0}_R$.  
Similarly, when the field theory is considered on the left wedge, the vacuum can be defined in the same way.

\vskip\baselineskip

\begin{figure}[h]
\centering
\includegraphics[keepaspectratio, scale=0.4]{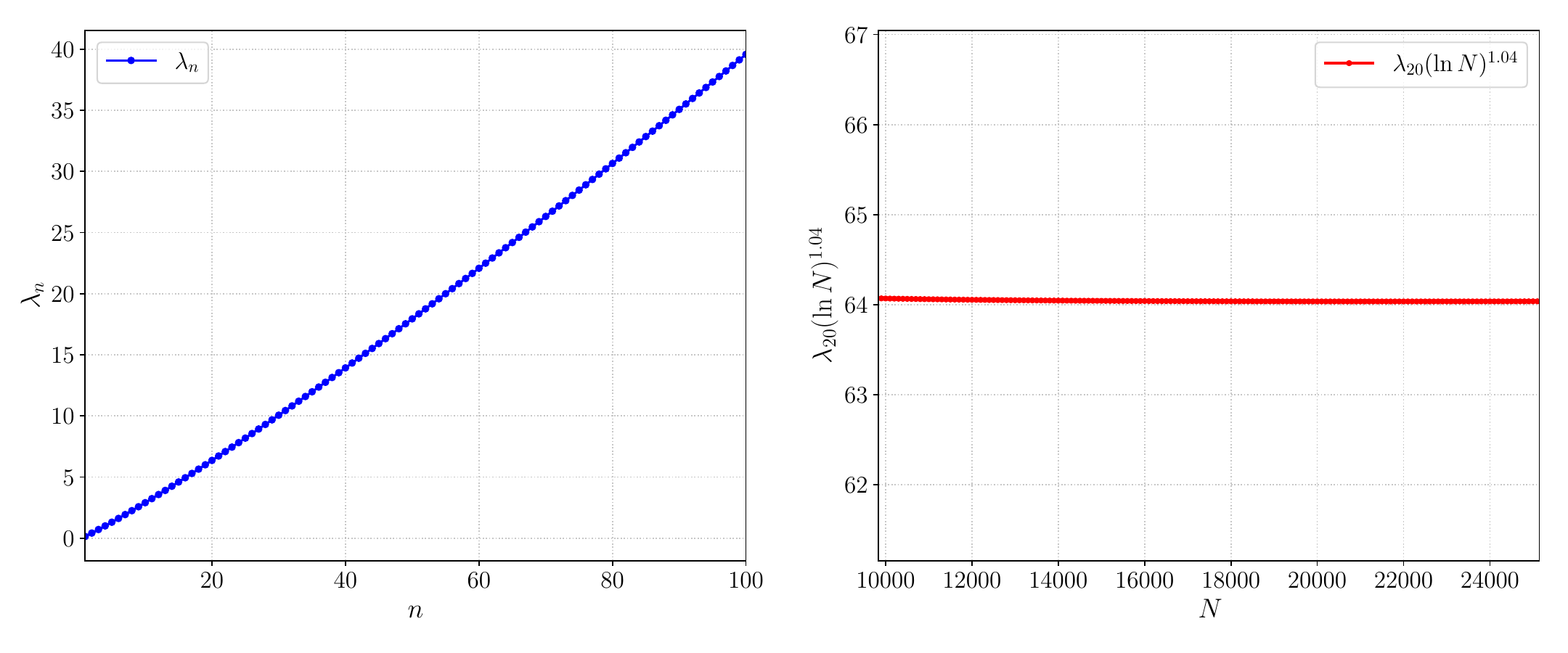}
\caption{$n$-dependence and $N$-dependence of $\lambda_n$.}
\label{fig8}
\end{figure}

Let us numerically compute the eigenvalues $\lambda_n$ of $A_{m,n}$ and examine their dependence on $n$ and $N$. From the left panel of Fig.~\ref{fig8}, we find that $\lambda_n$ is approximately linear in the energy level $n$. From the right panel of Fig.~\ref{fig8}, the dependence on the IR cutoff $N$ is found to be $\lambda_n \propto \frac{1}{(\ln N)^\gamma}$ with $\gamma \sim 1$.
In the left panel of Fig.~\ref{fig8}, $N$ is fixed to $10000$ while $n$ is varied as $n=1,2,\cdots,100$. In the right panel, $n$ is fixed to $20$ while $N$ is varied as $N=10000,10001,\cdots,25000$. In summary, for sufficiently large $N$ we find
\begin{align}
\lambda_n \sim \frac{n}{(\ln N)^\gamma},\qquad (\gamma \sim 1).
\label{sp}
\end{align}

This is consistent with the interpretation of our model as the lattice regularization of the free massless scalar in the Rindler wedge, because the metric of Rindler spacetime can be written as
\begin{align}
\dd{s^2}=e^{2 \kappa \xi} \left(-\dd{t^2}+\dd{\xi^2}
\right),
\end{align}
where
\begin{align}
x=\frac{1}{\kappa} e^{\kappa \xi},
\end{align}
and then the theory is equivalent to the theory on $\dd{s^2}=-\dd{t^2}+\dd{\xi^2}$ for the massless scalar field.
The space-like size is $\ln (x_N/x_1)/\kappa \sim \ln N/\kappa $
and the spectrum is consistent with \eqref{sp}, even though the lattice is not equal spacing in $\xi$.
Note that the light spectrum \eqref{sp} for $n \ll \ln N$ will correspond to the light modes of the brick wall model \cite{tHooft:1984kcu}, even though our model does not distinguish the light modes from the IR cut-off and the near-horizon region.
In the black-hole case, the far-IR region is no longer independent of the geometry, because the spacetime curvature becomes important at distances of order \(x \sim \kappa^{-1}\). It is therefore natural to identify the effective IR cutoff with this scale. On the lattice, this means that the system size \(N\) is not an independent parameter, but is set by
\[
N \sim \frac{\alpha}{\kappa a},
\]
where \(\alpha\) is an \({\cal O}(1)\) constant that depends on the precise definition of the cutoff. As a result,
\[
\ln N = \ln \alpha - \ln(\kappa a),
\]
and hence, up to an additive \({\cal O}(1)\) constant,
\[
\ln N \simeq -\ln(\kappa a).
\]
In this sense, the logarithmic term is effectively controlled by the UV quantity \(\kappa a\), rather than by an independent IR cutoff.

\paragraph{Remarks on the brick wall structure}
The ultraviolet (UV) regularization $1/a$ in Minkowski space implies a truncation of degrees of freedom within a distance $a/2$ from the horizon in Rindler coordinates (noting that $\dd{s}^2 = \dd{x}^2$ at $t = 0$). This effectively introduces a brick wall structure near $x \approx a$. Historically, the brick wall was introduced \cite{tHooft:1984kcu} to ensure that the black hole entropy remained finite, even if including the degrees of freedom of quantum fluctuations; this was achieved by discarding degrees of freedom within roughly a Planck length of the horizon. By identifying this Planck length with our UV cutoff $a$, our model directly corresponds to the brick wall framework. We note, however, that unlike the original brick wall model—which simply imposes a Dirichlet boundary condition—our approach also discretizes space to implement the UV regularization.

\subsection{Energy Density of the Rindler vacuum} \label{2.2}
In the continuum analysis of a scalar field in Rindler spacetime, let $T_{00}$ 
%$T^0_0=T_{00}$
denotes the energy density of the scalar field defined in Minkowski spacetime, let $\ket{0}_\mathrm{R}\equiv \ket{0}_\mathrm{R}^\mathrm{left}\otimes\ket{0}_\mathrm{R}^\mathrm{right}$ denotes the vacuum defined in the Rindler wedges, and let $\ket{0}_\mathrm{M}$ denote the vacuum defined in Minkowski spacetime. It is known that\footnote{In \cite{Parentani:1993yz}, it is stated that $\ev{T_{UU}}=-\frac{1}{48\pi}\frac{1}{x^2}$ in light-cone coordinates $U=T-X, V=T+X$, but the energy density is $\ev{T_{00}}=\ev{T_{UU}}+\ev{T_{VV}}=-\frac{1}{24\pi}\frac{1}{x^2}$.}
\begin{align}
\ev{T_{00}}\equiv{}_\mathrm{R}\mel{0}{T_{00}}{0}_\mathrm{R}-{}_\mathrm{M}\mel{0}{T_{00}}{0}_\mathrm{M}=-\frac{1}{24\pi}\frac{1}{x^2} \qquad (x>0).
\label{T_00}
\end{align}
The reason for subtracting the Minkowski vacuum expectation value is to remove the term that exhibits a UV divergence when taking the $a\to 0$ limit. Integrating \eqref{T_00} over the interval $0<x<\infty$ yields negative infinity, which appears to imply that a state with positive energy cannot be prepared, seemingly leading to a contradiction. However, this occurs because the contribution exactly on the horizon $x=0$ is not taken into account. Once it is included, the total energy becomes positive. At that point, the value of $\ev{T_{00}}$ exactly on the horizon is expressed as $\ev{T_{00}}\propto\frac{1}{\epsilon^2}$ using an infinitesimal positive number $\epsilon$, and diverges to positive infinity. (In detail about these results, see \cite{Parentani:1993yz, Czech:2012be, Germani:2015tda}.)
Such a divergence of the energy density on the horizon naively suggests that the vicinity of the horizon is a UV-sensitive region. Therefore, in this section we calculate $\ev{T_{00}}$ for the scalar field on lattice Rindler spacetime, introducing the UV cutoff $a$ defined in subsection \ref{2.1}. We will confirm that $\ev{T_{00}}>0$ appears only in the region near the horizon and that it indeed diverges to positive infinity exactly on the horizon as $a\to 0$. This demonstrates clearly that the near-horizon region is UV sensitive.

\vskip\baselineskip

Let us first define the energy density $T_{00}$ of the scalar field in Minkowski spacetime. The action is
\begin{align}
S&=-\frac{1}{2}\int\dd^2{X}\eta^{\mu\nu}\partial_\mu\phi\partial_\nu\phi \nonumber \\
&=\int\dd^2{X}\left[\frac{1}{2}(\partial_T\phi)^2-\frac{1}{2}(\partial_X\phi)^2\right]\equiv\int\dd^2{X}\mathcal{L}_\mathrm{M}.
\end{align}
Following the standard canonical formalism procedure, the energy density $T_{00}$ is the Hamiltonian density $\mathcal{H}_\mathrm{M}$, given by
\begin{align}
T_{00}=\mathcal{H}_\mathrm{M}=\frac{1}{2}(\partial_T\phi)^2+\frac{1}{2}(\partial_X\phi)^2.
\label{T_00_M}
\end{align}
Below, we focus only on the value of $T_{00}$ on the $T=0$ surface (which corresponds to the $t=0$ surface in Rindler spacetime). Noting the relations
\begin{align}
&\pdv{X}=-\frac{1}{\kappa x}\sinh\kappa t\pdv{t}+\cosh\kappa t\pdv{x}, \\
&\pdv{T}=\frac{1}{\kappa x}\cosh\kappa t\pdv{t}-\sinh\kappa t\pdv{x},
\end{align}
and rewriting $T_{00}\left.\right|_{t=0}$ in terms of the Rindler coordinates $\{t, x\}$, we obtain
\begin{align}
T_{00}&=\frac{1}{2}\frac{1}{(\kappa x)^2}(\partial_t\phi)^2+\frac{1}{2}(\partial_x\phi)^2 \nonumber \\
&=\frac{1}{2}\pi^2+\frac{1}{2}(\partial_x\phi)^2.
\label{T_00_Rindler}
\end{align}
Thus, we only need to discretize this expression. Incidentally, the above discussion can be applied in the same way to the left wedge as well as the right wedge. That is, for the scalar field $\phi_\mathrm{L}$ and its conjugate momentum $\pi_\mathrm{L}$ defined in the left wedge, $T_{00}\left.\right|_{t_\mathrm{L}=0}=\frac{1}{2}\pi_\mathrm{L}^2+\frac{1}{2}(\partial_{x_\mathrm{L}}\phi_\mathrm{L})^2$ holds on the $t_\mathrm{L}=0$ surface. Therefore, we can express $T_{00}$ in Rindler coordinates over the entire $T=0$ surface.
From \eqref{coordinateL} and \eqref{coordinateR}, noting that the coordinates $x_\mathrm{L/R}$ and $X$ coincide on the $t_\mathrm{L/R}=0$ surfaces, the discretized position on the $T=0$ surface can be written as
\begin{align}
X=
\begin{cases}
x_\mathrm{R}&(X>0)\\
-x_\mathrm{L}&(X<0)
\end{cases}
=\left(n-\frac{1}{2}\right)a\equiv X_n.
\end{align}
Here $n=1,2,3\cdots$ corresponds to the right wedge spacetime, while $n=0,-1,-2,\cdots$ corresponds to the left wedge. From the above, the discretized quantity $T_{00}(T=0, X=X_n)\equiv T_{n}$ can be written as
\begin{align}
T_n=\frac{1}{2}\pi_n^2+\frac{1}{2}\left[\frac{(\phi_{n+1}-\phi_{n})^2}{2a^2}+\frac{(\phi_{n}-\phi_{n-1})^2}{2a^2}\right].
\label{T_00_lattice}
\end{align}
As a reminder, $(\phi_n, \pi_n)$ are the scalar field and its conjugate momentum at $T=0, X=X_n$, defined on the right wedge for $n\geq 1$ and on the left wedge for $n\leq 0$. Moreover, when replacing $(\partial_x\phi)^2$ with a finite difference, we use a symmetric difference form so that the left and right wedges are treated equally.

\vskip\baselineskip

Next we compute ${}_\mathrm{R}\mel{0}{T_{n}}{0}_\mathrm{R}$ using \eqref{T_00_lattice}. Since the left and right wedges are symmetric under a $\pi$ rotation around the origin, it suffices to perform the calculation only in the right wedge. In the right wedge, using the relations
\begin{align}
&\pi_n=\frac{1}{\sqrt{n-\frac{1}{2}}}\sum_{m=1}^N (P^T)_{n,m}\sqrt{\frac{\lambda_m}{2a^2}}(a_m^\dag+a_m), \label{Bogo1}\\
&\phi_n=\sqrt{n-\frac{1}{2}}\sum_{m=1}^N (P^T)_{n,m}\frac{-i}{\sqrt{2\lambda_m}}(a_m^\dag-a_m), \label{Bogo2}
\end{align}
and noting that for $n\geq 2$ the operators $a_m^\dag, a_m$ act only on the right states, we obtain
\begin{align}
{}_\mathrm{R}\mel{0}{T_n}{0}_\mathrm{R}&=\left({}^\mathrm{left}_\mathrm{R}\bra{0}\otimes{}^\mathrm{right}_\mathrm{R}\bra{0}\right) T_{n}\left(\ket{0}_\mathrm{R}^\mathrm{left}\otimes\ket{0}_\mathrm{R}^\mathrm{right}\right) \nonumber \\
&={}^\mathrm{right}_\mathrm{R}\mel{0}{T_{n}}{0}^\mathrm{right}_\mathrm{R} \nonumber \\
&=\frac{1}{4a^2\left(n - \frac{1}{2}\right)} (A^{1/2})_{nn} + \frac{n + \frac{1}{2}}{8a^2} (A^{-1/2})_{n+1, n+1}+ \frac{n - \frac{1}{2}}{4a^2} (A^{-1/2})_{nn}  \notag \\
&\qquad + \frac{n - \frac{3}{2}}{8a^2} (A^{-1/2})_{n-1, n-1}- \frac{\sqrt{\left(n+\frac{1}{2}\right)\left(n-\frac{1}{2}\right)}}{4a^2} (A^{-1/2})_{n, n+1} \nonumber \\
&\qquad - \frac{\sqrt{\left(n-\frac{1}{2}\right)\left(n-\frac{3}{2}\right)}}{4a^2} (A^{-1/2})_{n-1, n}.
\label{T_n_R}
\end{align}
During the calculation, we used the fact that for a function $f(\lambda_n)$ of $\lambda_n$,
\begin{align}
\sum_{j=1}^N (P^T)_{n,j}(P^T)_{m,j}f(\lambda_j)&=\sum_{j=1}^N P_{j,n}f(\lambda_j)(P^T)_{m,j} \nonumber \\
&=[f(A)]_{n,m}.
\end{align}
On the other hand, special care is required for the case $n=1$. The $\phi_1\phi_0$ term appearing in \eqref{T_00_lattice} has $\phi_1$ acting on the right wedge state and $\phi_0$ acting on the left wedge state, which leads to
\begin{align}
{}_\mathrm{R}\mel{0}{\phi_1\phi_0}{0}_\mathrm{R}={}^\mathrm{right}_\mathrm{R}\mel{0}{\phi_1}{0}^\mathrm{right}_\mathrm{R}\cdot {}^\mathrm{left}_\mathrm{R}\mel{0}{\phi_0}{0}^\mathrm{left}_\mathrm{R}=0.
\end{align}
Furthermore, from the symmetry between the right and left wedges, the expectation value of the $\phi_0^2$ term can be evaluated as
\begin{align}
{}_\mathrm{R}\mel{0}{\phi_0^2}{0}_\mathrm{R}&={}^\mathrm{left}_\mathrm{R}\mel{0}{\phi_0^2}{0}^\mathrm{left}_\mathrm{R} \nonumber \\
&={}^\mathrm{right}_\mathrm{R}\mel{0}{\phi_1^2}{0}^\mathrm{right}_\mathrm{R},
\end{align}
and therefore
\begin{align}
{}_\mathrm{R}\mel{0}{T_{n=1}}{0}_\mathrm{R}&=\frac{1}{2} (A^{1/2})_{1,1} + \frac{3}{16} (A^{-1/2})_{2,2} + \frac{3}{16} (A^{-1/2})_{1,1} - \frac{\sqrt{3}}{8} (A^{-1/2})_{1,2}.
\end{align}

\vskip\baselineskip
Next, we determine ${}_\mathrm{M}\mel{0}{T_{00}}{0}_\mathrm{M}$. Discretizing with a lattice cut $X=\left(n-\frac{1}{2}\right)a=X_n\,\,\,(n=0,1,\cdots , N)$, and using the notation $\phi(X=X_n)=\phi_n$ and $\pi(X=X_n)=\pi_n$ as before, the Hamiltonian density and Hamiltonian are written as
\begin{align}
\mathcal{H}_\mathrm{M}=\frac{1}{2}\pi_n^2+\frac{1}{2a^2}(\phi_{n+1}-\phi_{n})^2,
\end{align}
\begin{align}
H_\mathrm{M}&=a\sum_{n=1}^{N} \left(\frac{1}{2}\pi_n^2+\frac{1}{2a^2}(\phi_{n+1}-\phi_{n})^2\right) \label{Hamiltonian_M1}\\
&=a\sum_{n=1}^{N} \left(\frac{1}{2}\pi_n^2+\frac{1}{2a^2}\sum_{m=1}^N\phi_mB_{m,n}\phi_n\right), \label{Hamiltonian_M2} \\
B&=
\mqty(
1 & -1 & 0 & \cdots & \cdots & 0 \\
-1 & 2 & -1 & 0 & \cdots & 0 \\
0 & -1 & 2 & -1 & \ddots & \vdots \\
\vdots & 0 & \ddots & \ddots & \ddots & 0 \\
\vdots & \vdots & \ddots & -1 & 2 & -1 \\
0 & 0 & \cdots & 0 & -1 & 1
).
\end{align}
Here, $\phi_{N+1}$, which appears due to replacing derivatives with finite differences in \eqref{Hamiltonian_M1} and \eqref{Hamiltonian_M2}, is eliminated by imposing the boundary condition $\phi_{N+1}=0$.  
Using an orthogonal matrix $Q$ such that $QBQ^T=\mathrm{diag}(\Lambda_n^2)$ and defining
\begin{align}
\Pi_n^\mathrm{M}=a \sum_{m=1}^N Q_{n,m}{\pi}_m, \qquad
\Phi_n^\mathrm{M}=\sum_{m=1}^N Q_{n,m}{\phi}_m,
\label{Bogo3}
\end{align}
\eqref{Hamiltonian_M2} becomes
\begin{align}
H_\mathrm{M}=\frac{1}{2 a}\sum_{n=1}^{N} \left((\Pi_n^\mathrm{M})^2+\Lambda_n^2 (\Phi_n^\mathrm{M})^2\right).
\end{align}
Imposing the canonical commutation relation $[\phi_m, \pi_n]=i\delta_{m,n}/a$, we can introduce the creation and annihilation operators
\begin{align}
b_n^\dag=\frac{1}{\sqrt{2\Lambda_n}}\left(\Pi_n^\mathrm{M}+i \Lambda_n \Phi_n^\mathrm{M}\right),\qquad
b_n=\frac{1}{\sqrt{2\Lambda_n}}\left(\Pi_n^\mathrm{M}-i \Lambda_n \Phi_n^\mathrm{M}\right),
\label{Bogo4}
\end{align}
to rewrite the Hamiltonian as
\begin{align}
H_\mathrm{M}=\sum_{n=1}^{N} \frac{\Lambda_n}{a}\left(b_n^\dag b_n+\frac{1}{2}\right).
\end{align}
These operators $b_n^\dag, b_n$ satisfy the commutation relation
\begin{align}
[b_m, b_n^\dag]=\delta_{m,n},
\end{align}
and the vacuum $\ket{0}_\mathrm{M}$ in Minkowski spacetime is defined by $b_n\ket{0}_\mathrm{M}=0\,\,\,(^\forall n)$.  
As in the Rindler spacetime case, ${}_\mathrm{M}\mel{0}{T_{00}}{0}_\mathrm{M}$ can be calculated by expressing $\phi_n, \pi_n$ in terms of $b_n^\dag, b_n$. The result is
\begin{align}
{}_\mathrm{M}\mel{0}{T_{00}}{0}_\mathrm{M}
&=\frac{1}{4a^2}(B^{1/2})_{n,n}
+\frac{1}{16a^2}(B^{-1/2})_{n+1,n+1}\nonumber\\
&\qquad +\frac{1}{16a^2}(B^{-1/2})_{n-1,n-1}
-\frac{1}{8a^2}(B^{-1/2})_{n+1,n-1}.
\label{T_n_M}
\end{align}

Recall that we compute ${}_\mathrm{M}\mel{0}{T_{00}}{0}_\mathrm{M}$ in order to subtract the UV-divergent contribution appearing in ${}_\mathrm{R}\mel{0}{T_{00}}{0}_\mathrm{R}$ in the limit $a\to 0$. Although we have just calculated ${}_\mathrm{M}\mel{0}{T_{00}}{0}_\mathrm{M}$ within the lattice theory, its value is strongly affected by boundary effects near $n=1, N$, and therefore differs from the value we actually wish to obtain. 
Moreover, the quantity ${}_\mathrm{R}\mel{0}{T_{00}}{0}_\mathrm{R}$ computed in Rindler spacetime is also affected by the artificial boundary at $n=N$, and therefore deviates from its intrinsic value.

The IR cutoff associated with the asymptotic region of Minkowski space is not the main issue we wish to focus on here. We therefore restrict attention to a region with \(n \ll N\), while still keeping the region sufficiently large. In this regime, the energy density in the Minkowski vacuum should be approximately translationally invariant. For this reason, and to simplify the calculation, it is reasonable to use the energy density ${}_\mathrm{M}\mel{0}{T_{00}}{0}_\mathrm{M}$ evaluated near the middle of the \(N\)-site Minkowski Hamiltonian, namely at \(n = N/2\), and we define $\ev{T_{00}}$ as ${}_\mathrm{R}\mel{0}{T_{00}}{0}_\mathrm{R}$ minus this value.

%To address this issue, for ${}_\mathrm{M}\mel{0}{T_{00}}{0}_\mathrm{M}$ we adopt the following prescription: we regard the value of ${}_\mathrm{M}\mel{0}{T_{00}}{0}_\mathrm{M}$ at $n=N/2$, which is farthest from the boundaries, as the true bulk value of ${}_\mathrm{M}\mel{0}{T_{00}}{0}_\mathrm{M}$, and we define $\ev{T_{00}}$ as ${}_\mathrm{R}\mel{0}{T_{00}}{0}_\mathrm{R}$ minus this value. In fact, since the boundary effect at a distance $r$ from the boundary scales as $\sim \frac{1}{r^2}$, taking a sufficiently large $N$ ensures adequate numerical accuracy. The reason why the value at $n=N/2$ can be taken as representative of ${}_\mathrm{M}\mel{0}{T_{00}}{0}_\mathrm{M}$ is that this quantity is defined in Minkowski spacetime and is therefore independent of position.

%On the other hand, it is difficult to estimate and correct the deviation of ${}_\mathrm{R}\mel{0}{T_{00}}{0}_\mathrm{R}$ near $n=N$. Therefore, in what follows, we restrict our attention to the region $n \ll N$, where the boundary effects are negligible.

\vskip\baselineskip

\begin{figure}[h]
\centering
\includegraphics[keepaspectratio, scale=0.4]{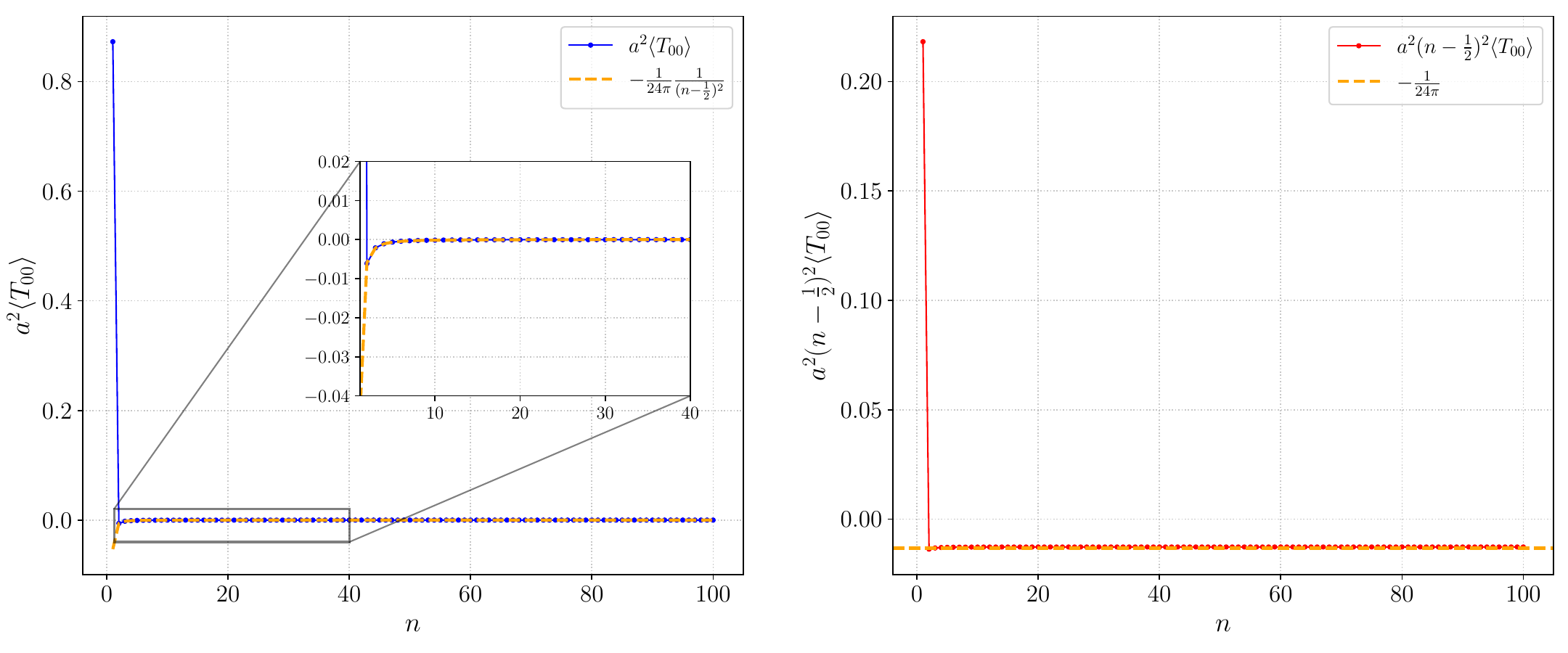}
\caption{Plots of $a^2\ev{T_{00}}$ (left) and $a^2(n-\frac{1}{2})^2\ev{T_{00}}$ (right) for $N=10000$. The yellow lines represent the continuum theory.}
\label{fig1}
\end{figure}

Setting $N=10^4$ and numerically evaluating $\ev{T_{00}}$ on the $T=0$ surface from \eqref{T_n_R} and \eqref{T_n_M} yields the results shown in Figure \ref{fig1}. From the figure, we can see that the energy density, which diverges exactly on the horizon in the continuum theory, leaks out to $n=1$ outside the horizon due to the cutoff effect.  
The results for $n\geq 2$ are consistent with the continuum expression \eqref{T_00}, and from the right panel we see that the coefficient $\frac{1}{24\pi}$ is correctly reproduced. Moreover, because $a^2\ev{T_{00}}(n=1)\simeq 0.87$ is significantly positive, the total energy from $n=1$ to $n=100$, $a^2\sum_{n=1}^{100} \ev{T_{00}}(n)\simeq 0.86$,
does not become negative. Since the value of $\ev{T_{00}}$ at $n=100$ is $-\mathcal{O}(10^{-6})$, we can expect that $\sum_{n=1}^{N} \ev{T_{00}}(n)>0$ holds even for sufficiently large $N$.  
This indicates that 
the lattice makes explicit how the horizon-supported positive contribution is smeared over the stretched-horizon region.
%the negative-energy problem and the divergence of the energy density on the horizon that appear in the continuum theory can be resolved by introducing a cutoff as a regularization.
Note also that not just the energy density at the horizon, but also the total energy, diverges in the continuum theory.

The explicit values of $a^2\ev{T_{00}}(n)$ are listed in Table~\ref{tabT_00}. We also examine whether $a^2\sum_{n=1}^{100}\ev{T_{00}}(n)$ depends on the IR cutoff $N$. The values of $a^2\sum_{n=1}^{100}\ev{T_{00}}(n)$ for different choices of $N$ are listed in Table~\ref{tab_sumT_00}. Although a mild dependence on $N$ is observed, the variation tends to stabilize as $N$ becomes larger.

\begin{table}[h]
  \centering
  \caption{Values of $a^2\ev{T_{00}}(n)$}
  \begin{tabular}{|c|c|}
    \hline
    $n$ & $a^2\ev{T_{00}}(n)$ \\ \hline
    1 & $8.72804176 \times 10^{-1}$ \\
    2 & $-6.10776159 \times 10^{-3}$ \\
    3 & $-2.11302198 \times 10^{-3}$ \\
    4 & $-1.06189513 \times 10^{-3}$ \\
    5 & $-6.37731106 \times 10^{-4}$ \\ \hline
  \end{tabular}
\label{tabT_00}
\end{table}

\begin{table}[h]
  \centering
  \caption{Values of $a^2\sum_{n=1}^{100}\ev{T_{00}}(n)$}
  \begin{tabular}{|c|c|}
    \hline
    $N$ & $a^2\sum_{n=1}^{100}\ev{T_{00}}(n)$ \\ \hline
    10000 & $8.6046147064 \times 10^{-1}$ \\
    12000 & $8.6297995109 \times 10^{-1}$ \\
    14000 & $8.6507940032 \times 10^{-1}$ \\
    16000 & $8.6687643872 \times 10^{-1}$ \\
    18000 & $8.6844526104 \times 10^{-1}$ \\
    20000 & $8.6983593325 \times 10^{-1}$ \\ \hline
  \end{tabular}
\label{tab_sumT_00}
\end{table}

\subsubsection{Other factorized states}

Let us consider a pure state in a Rindler wedge $\ket{s}$, instead of the Rindler vacuum and the tensor product of them
as 
\begin{align}
\ket{s}_\mathrm{R} =
\ket{s}_\mathrm{R}^\mathrm{left}\otimes\ket{s}_\mathrm{R}^\mathrm{right}. 
\end{align}
Below, we will show that this state should be drastically different from the Minkowski vacuum at the brick wall (i.e., $n=1$), as in the Rindler vacuum.
We will concentrate on the operator $\pi_0 \pi_1$.%, which is contained in $T_1$.
The expectation value of this is given by 
${}_M \mel{0}{\pi_0 \pi_1}{0}_M=\frac{c}{a^2} $, where $c$ is a ${\cal O}(1)$ constant, for the Minkowski vacuum, and 
${}_\mathrm{R} \mel{s}{\pi_0 \pi_1}{s}_\mathrm{R}=({}_\mathrm{R} \mel{s}{\pi_1}{s}_\mathrm{R})^2$ for the factorized state $\ket{s}_\mathrm{R}$.
Thus, the difference ${}_\mathrm{R} \mel{s}{\pi_0 \pi_1}{s}_\mathrm{R}-{}_M \mel{0}{\pi_0 \pi_1}{0}_M={\cal O}(1/a^2)$, except ${}_\mathrm{R} \mel{s}{\pi_1}{s}_\mathrm{R} =\frac{\sqrt{c}}{a}$ is satisfied very precisely.
This means that the difference is very large, otherwise  ${}_\mathrm{R} \mel{s}{\pi_1}{s}_\mathrm{R}$ is very large. Here  ${}_\mathrm{M} \mel{0}{\pi_1}{0}_\mathrm{M}=0$.
Note that we can straightforwardly generalize this discussion for the tensor product of a pure state to the tensor product of a mixed state.
We can also generalize this discussion to the tensor product of different left and right states.
If one considers a theory with even very weak interactions, then such a large deviation from the Minkowski vacuum at the brick wall would prevent a wave packet in this state from simply passing through that region. Here, the time evolution is taken with respect to the Minkowski Hamiltonian. As a result, after a sufficiently long time, as in \eqref{st}, this effect becomes visible even when the horizon is still far away. In other words, once a cut-off is present, the Minkowski vacuum and a factorized state exhibit different behavior in the presence of interactions.
Note also that, if one starts from a pure state and specifies the state on the right Rindler wedge, then requiring the combined left-right state to remain pure after tracing out the left degrees of freedom forces the total state to be factorized.

\subsection{Retarded Green Function}\label{2.3}
Introducing a UV cutoff in Rindler spacetime effectively creates a boundary of spacetime just outside the horizon. 
This boundary can be regarded as an object analogous to a brick wall\footnote{
Because there is no degree of freedom inside the $n=1$ site, we will call this object a brick wall, instead of the stretched horizon.
}, and in this subsection, we argue that this interpretation is indeed correct. The method is to compute the Green function of a massless free scalar field in Rindler spacetime and examine the response of the scalar field. A wave incident toward the horizon from outside would approach the horizon indefinitely (in Rindler time) if there were no cutoff; however, when the cutoff is present, it is expected to be reflected near $n=1$. By examining the time evolution of the Green function, we confirm the presence of this reflection.

\begin{figure}[h]
\centering
\begin{tikzpicture}[scale=0.7, ray/.style={thick, orange, postaction={decorate}, decoration={markings, mark=at position 0.6 with {\arrow{Stealth[length=3mm]}}}}]
\draw[-{Stealth[length=3mm]}, thick] (-0.5, 0) -- (6.5, 0) node[below] {$X$};
\draw[-{Stealth[length=3mm]}, thick] (0, -1.5) -- (0, 6.5) node[left] {$T$};

\draw[dashed, gray, thick] (0,0) -- (6,6) node[pos=0.6, above left, inner sep=4pt] {horizon};
\draw[dashed, gray, thick] (0,0) -- (1.5,-1.5);

\draw[gray, thin, densely dashed] (5, -1.5) -- (5, 6);

\draw[ultra thick, blue, domain=-1.5:5.8, smooth, variable=\t] plot ({sqrt(1+\t*\t)}, {\t});

\node[blue, right=4pt] at ({sqrt(1+(5.2)*(5.2))}, 5.2) {stretched horizon};

\coordinate (Start) at (3, 0);
\coordinate (Bounce) at (5/3, 4/3);
\coordinate (End1) at (5, 14/3);
\coordinate (End2) at (5, 2);

\draw[ray](Start)--(Bounce);
\draw[ray](Bounce)--(End1);
\draw[ray](Start)--(End2);

\filldraw[black] (Start) circle (2pt);
\node[below=3pt] at (Start) {$X=X_1$};

\filldraw[black] (End1) circle (2pt);
\node[below right=1pt] at (End1) {$X=X_2$};

\filldraw[black] (End2) circle (2pt);

\filldraw[red] (Bounce) circle (2pt);

\end{tikzpicture}
\caption{Illustration of a field reflected by the stretched horizon.}
\label{fig.image}
\end{figure}
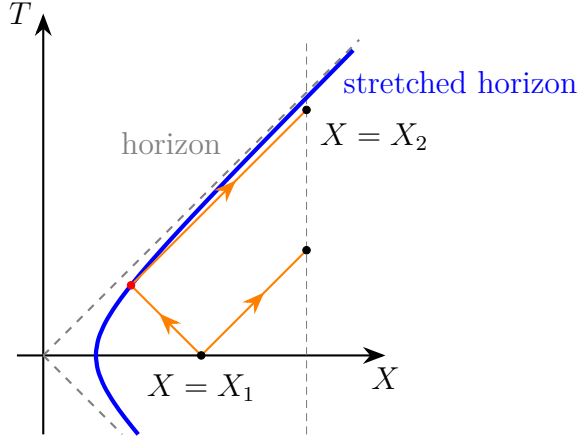

\begin{figure}[h]
\centering
\includegraphics[keepaspectratio, scale=0.45]{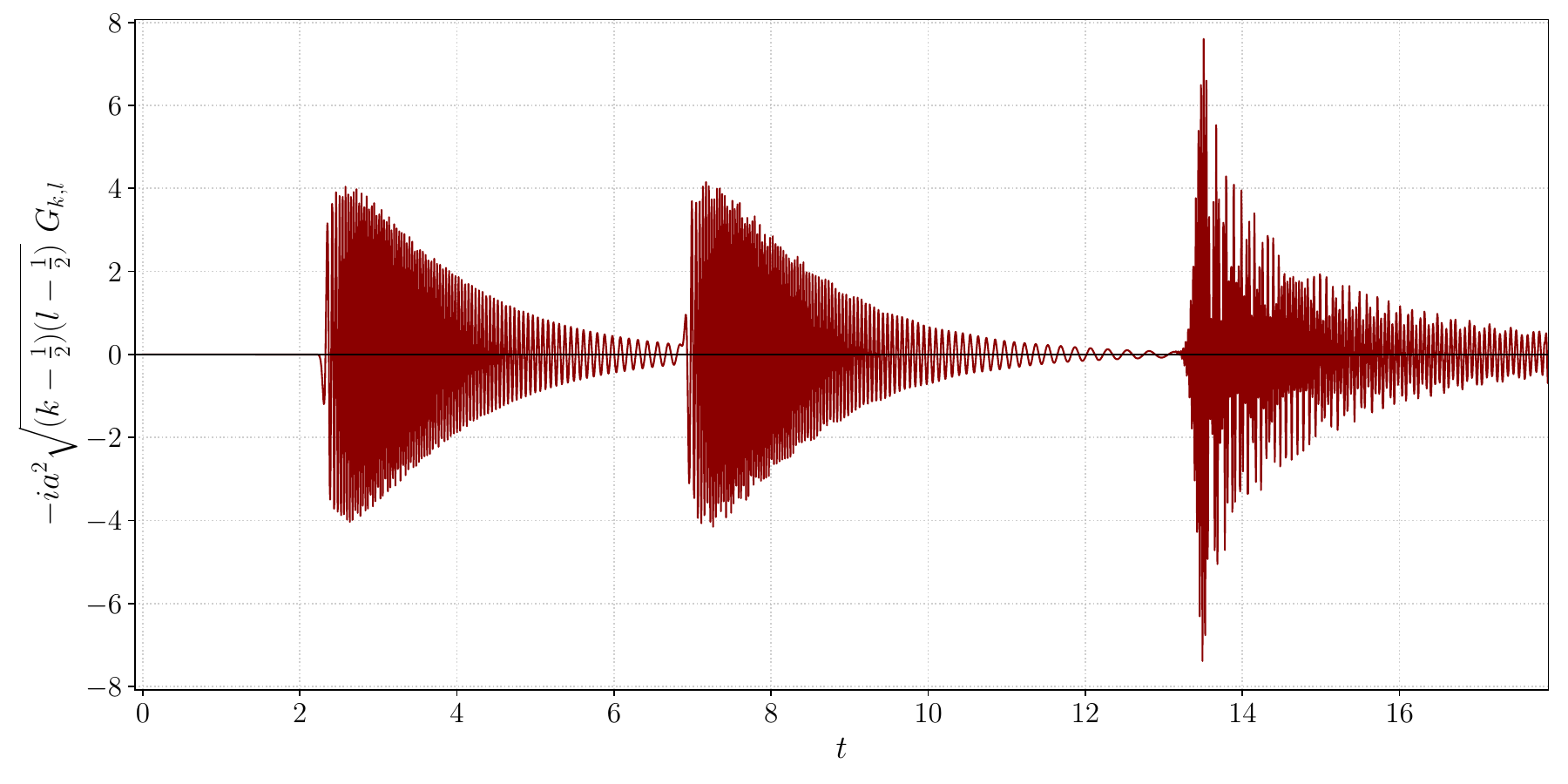}
\caption{Plot of $-ia^2\sqrt{(j_1-\frac{1}{2})(j_2-\frac{1}{2})}\,\,G_{j_1,j_2}$.\\
We set $N=10000, j_1=100, j_2=1000$, and normalize to $\kappa=1$.}
\label{fig2}
\end{figure}

\vskip\baselineskip
The Green function in Rindler spacetime is given by ${}_\mathrm{R}\mel{0}{[\phi(0,x), \phi(t,y)]}{0}_\mathrm{R}$. However, in this expression, the contribution of the zero mode of $\phi$ becomes dominant, which makes the response difficult to observe numerically. Therefore, we instead compute the vacuum expectation value of the commutator of the canonical momentum, ${}_\mathrm{R}\mel{0}{[\pi(0,x), \pi(t,y)]}{0}_\mathrm{R}$. Since $\pi=\partial_t\phi$, the contribution from the zero mode is absent.
%low-energy modes is suppressed. 
Taking $t>0$ and defining $x=a(j_1-1/2), \,y=a(j_2-1/2)$, we obtain
\begin{align}
{}_\mathrm{R}\mel{0}{[\pi(0,x), \pi(t,y)]}{0}_\mathrm{R}&={}_\mathrm{R}\mel{0}{[\pi_{j_1}, e^{iHt}\,\pi_{j_2}\,e^{-iHt}]}{0}_\mathrm{R} \\
&={}_\mathrm{R}\mel{0}{\pi_{j_1}\,e^{iHt}\pi_{j_2}\,e^{-iHt}}{0}_\mathrm{R}-{}_\mathrm{R}\mel{0}{e^{iHt}\,\pi_{j_2}\,e^{-iHt}\pi_{j_1}}{0}_\mathrm{R} \\
&=\frac{i}{a^2}\frac{1}{\sqrt{(j_1-\frac{1}{2})(j_2-\frac{1}{2})}}\sum_{m=1}^N (P^T)_{j_1,m}(P^T)_{j_2,m}\lambda_m\sin\kappa\lambda_m t\equiv G_{j_1,j_2}.
\label{green}
\end{align}
In the intermediate steps, we used
\begin{align}
&\pi_n=\frac{1}{\sqrt{n-\frac{1}{2}}}\sum_{m=1}^N (P^T)_{n,m}\sqrt{\frac{\lambda_m}{2a^2}}(a_m^\dag+a_m), \\
&a^\dag_m\,e^{-iHt}=e^{-iHt}\cdot e^{i\kappa\lambda_m t}\,a_m^\dag,  \\
&a_m\,e^{-iHt}=e^{-iHt}\cdot e^{-i\kappa\lambda_m t}\,a_m. 
\end{align}
Fixing $N=10000, \,\,j_1=100,\,\, j_2=1000$ and varying $t$, we compute Eq.\,\eqref{green} numerically. The result is shown in Fig.\,\ref{fig2}.

\noindent
To interpret this result, let us first consider the classical trajectory of a field propagating at the speed of light in Rindler spacetime. The field propagates along geodesics satisfying $\dd{s}^2=-\kappa^2 x^2\dd{t}^2+\dd{x}^2=0$, which implies
\begin{align}
\dd{(\kappa t)}=\pm\frac{\dd{(\kappa x)}}{\kappa x}.
\end{align}
Therefore, the time required for the field to travel from $x=x_1$ to $x=x_2$ is
\begin{align}
\Delta t=\frac{1}{\kappa}\ln\abs{\frac{x_2}{x_1}}. 
\end{align}
In the following, we set $\kappa=1$. The plot in Fig.\,\ref{fig2} contains three responses, which can be interpreted as follows.

\paragraph{\underline{The First Response}}
The first response can be interpreted as the field propagating directly from $x=x_{j_1}$ to $x=x_{j_2}$. The classical propagation time from $x_{j_1}$ to $x_{j_2}$ is $\Delta t=\ln\frac{1000-\frac{1}{2}}{100-\frac{1}{2}}\sim2.3$. Since the propagation starts at $t=0$, the response should appear at $t\sim2.3$, and indeed the response begins around that time.

\paragraph{\underline{The Second Response}}
The second response corresponds to the first signal reaching the boundary at $x=x_N$, being reflected there, and then returning to $x=x_{j_2}$. This is an artifact caused by making the spatial region finite. The time required for this process is $\Delta t=\ln\frac{1000-\frac{1}{2}}{100-\frac{1}{2}}+2\ln\frac{10000-\frac{1}{2}}{1000-\frac{1}{2}}\sim6.9$, which agrees with the starting time of the second response.

\paragraph{\underline{The Third Response}}
The third response can be interpreted as the field first propagating from $x=x_{j_1}$ toward the horizon, then being reflected by the boundary located near the horizon, and finally propagating to $x=x_{j_2}$. If the boundary is located at $x=x_1$, the time required for this process is $\Delta t=\ln\frac{100-\frac{1}{2}}{\frac{1}{2}}+\ln\frac{1000-\frac{1}{2}}{\frac{1}{2}}\sim12.9$, and the response indeed begins around that time.
\vskip\baselineskip
\noindent
Here, we have fixed the observation points to $j_1=100$ and $j_2=1000$, but similar reflections from the horizon and the boundary at $n=N$ can be observed for other choices of positions as well.

\vskip\baselineskip
In summary, when a UV cutoff is introduced in a theory defined on a spacetime with a horizon, a boundary effectively appears just outside the horizon. Since this boundary reflects fields propagating toward the horizon, the system effectively behaves as if there 
%were a stretched horizon or 
was a brick wall.
Note that the retarded Green function for the free field does not depend on the state.
Thus, this reflection from the brick wall exists for any state, including the canonical ensemble.

\subsubsection{Discussions on Hamiltonian in Rindler wedge}

We now discuss how the reflection at the brick wall depends on the choice of the Rindler Hamiltonian. We are interested in lattice Rindler Hamiltonians that reproduce the continuum theory in the limit \(a \to 0\). Although the Hamiltonian used in this paper may appear exceptional because of the brick wall, it is still consistent with the continuum limit, since the region where the brick-wall effect is visible vanishes as \(a \to 0\), as in \eqref{st}. We also note that, once one adopts a Rindler Hamiltonian, the description is restricted from the outset to a single Rindler wedge.

One may modify the Hamiltonian by adding terms built from fields near \(n=1\). Such terms should not obstruct the continuum limit, although in the continuum they amount to changing the condition imposed at the horizon as a lightlike boundary. However, terms involving \(n=0\) cannot be introduced by construction. As a result, unlike the Minkowski Hamiltonian, the Rindler Hamiltonian is closed entirely within the region \(n>0\), and hence inevitably leads to reflection at the brick wall.

This suggests that the singular dependence of the Rindler description on the UV cut-off near \(x=0\) makes reflection at the brick wall unavoidable.\footnote{Even in special cases where the exact modular Hamiltonian can be represented by a local corner Hamiltonian \cite{Okunishi:2019dmv}, the essential issue remains the same. If the Rindler-time evolution is implemented by a local lattice Hamiltonian acting only on the half-space, then there is no transmission channel across the horizon. In the absence of additional nonlocal couplings or dissipative degrees of freedom, a brick-wall-like reflection is therefore not a special feature of our particular construction, but a rather generic consequence of realizing the Rindler evolution at finite cutoff by a closed half-space lattice Hamiltonian.}
This statement is state-independent: it applies not only to pure states such as typical states, but also to the reduced state obtained by tracing out one side of the Minkowski vacuum. Thus, once a cut-off is introduced, the global description based on the Minkowski Hamiltonian and the wedge description based on the Rindler Hamiltonian become essentially different at the operator level. It would be very interesting to understand whether the same conclusion also holds in the black-hole case.

\subsection{Thermal Property}\label{2.4}
In the continuum theory, the Minkowski vacuum $\ket{0}_\mathrm{M}$ takes the form of a thermofield double (TFD) state of the left and right wedges\cite{Unruh:1976db, Israel:1976ur, Bisognano:1976za}:
\begin{align}
\ket{0}_\mathrm{M}=\frac{1}{\sqrt{Z}}\sum_{n} e^{-\pi E_n/\kappa}\,\ket{E_n}^\mathrm{left}\otimes\ket{E_n}^\mathrm{right}.
\label{TFD}
\end{align}
By tracing out the state in one wedge, a canonical ensemble is obtained, which implies that the Minkowski vacuum appears as a thermal state when viewed in Rindler spacetime. Indeed, using the creation and annihilation operators $a^\dag (\omega), a(\omega)$ of a massless scalar field defined in Rindler spacetime in the continuum theory, the expectation value of the particle number is
\begin{align}
{}_\mathrm{M}\mel{0}{a^\dag(\omega)a(\omega)}{0}_\mathrm{M}=\frac{1}{e^{2\pi\omega/\kappa}-1}, 
\end{align}
which follows the Bose distribution, and an observer in Rindler spacetime observes thermal radiation. (\cite{Fulling:1972md, Davies:1974th, Unruh:1976db, DeWitt:1975ys})

Does this thermal property of the Minkowski vacuum also appear in the lattice theory with a cutoff? In this section, we conclude that when a cutoff is introduced, $\ket{0}_\mathrm{M}$ deviates from a thermal state. In other words, in a theory with a cutoff, the Minkowski vacuum inevitably contains not only entanglement between states with the same energy levels in the left and right wedges, but also entanglement between different energy levels.

\vskip\baselineskip
To demonstrate this, using the creation and annihilation operators $a^\dag_n, a_n$ defined in Rindler spacetime and the Minkowski vacuum $\ket{0}_\mathrm{M}$ introduced in \ref{2.1} and \ref{2.2}, we compute
\begin{align}
N_{m,n}\equiv{}_\mathrm{M}\mel{0}{a^\dag_m a_n}{0}_\mathrm{M}. 
\label{N}
\end{align}
If $\ket{0}_\mathrm{M}$ were a TFD state even in the lattice formalism as in Eq.\,\eqref{TFD}, this quantity would exhibit a thermal distribution for $m=n$ and would vanish for $m\neq n$.

\begin{figure}[h]
\centering
\includegraphics[keepaspectratio, scale=0.4]{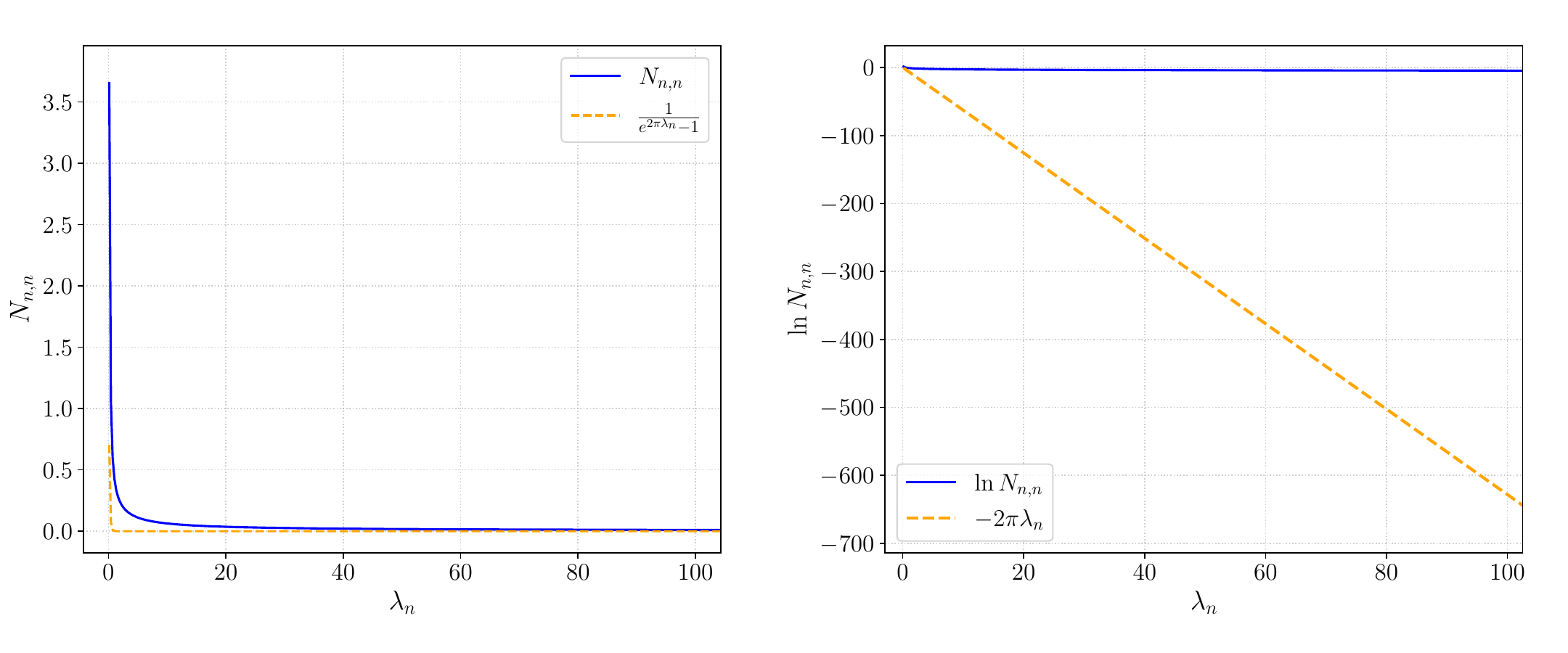}
\caption{Plots of $N_{n,n}$ (left) and $\ln N_{n,n}$ (right) for $N=10000$. The yellow line in the left figure is the Bose distribution, and the yellow line in the right figure is the Bose distribution for $\lambda_n\gg 1$.}
\label{fig3}
\end{figure}

\begin{figure}[h]
\centering
\includegraphics[keepaspectratio, scale=0.45]{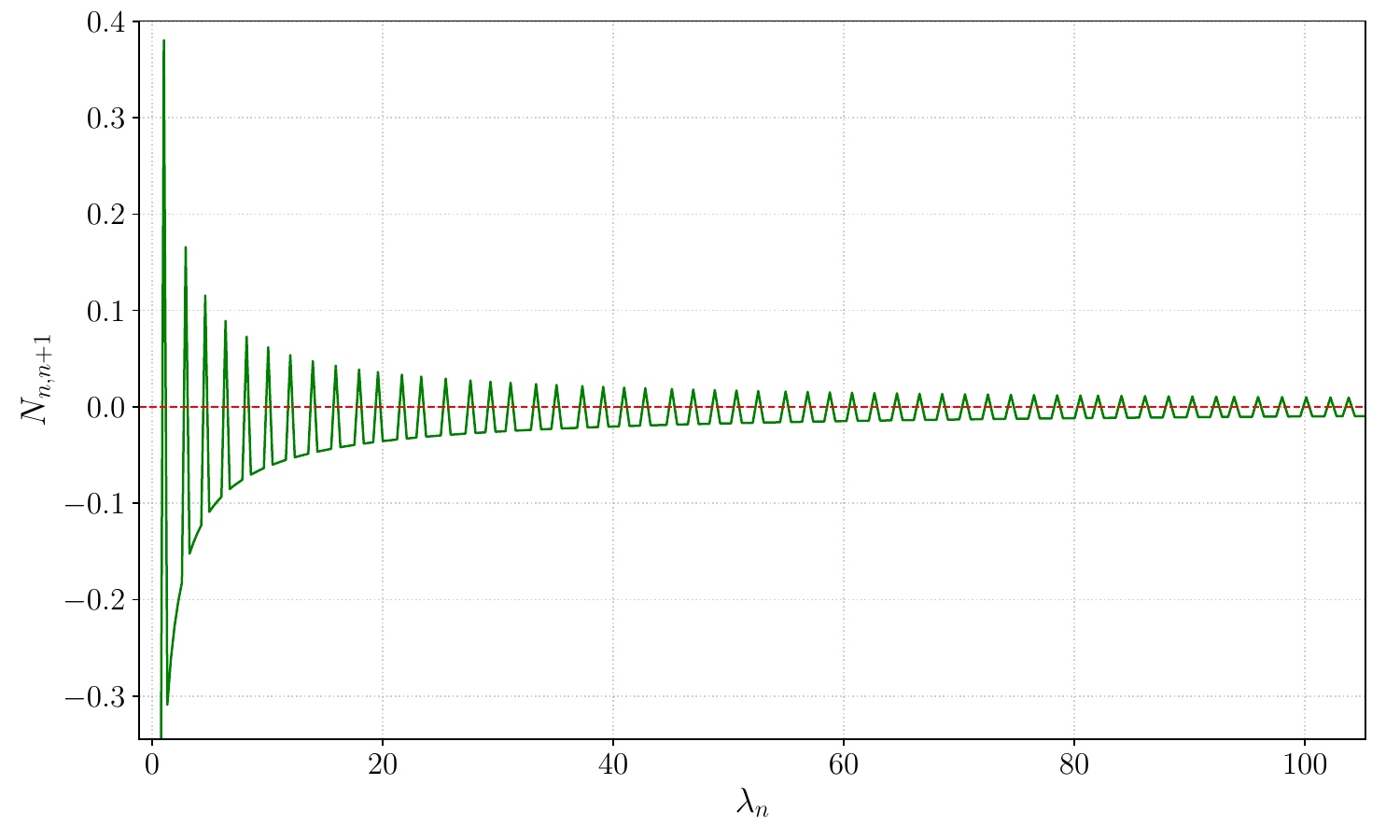}
\caption{Plot of $N_{n,n+1}$ calculated for $N=10000$.}
\label{fig4}
\end{figure}

To compute \eqref{N}, we use
\begin{align}
&\pi_n=\sum_{m=1}^N (Q^T)_{n,m}\sqrt{\frac{\Lambda_m}{2a^2}}(b_m^\dag+b_m), \label{Bogo5}\\
&\phi_n=\sum_{m=1}^N (Q^T)_{n,m}\frac{-i}{\sqrt{2\Lambda_m}}(b_m^\dag-b_m),  \label{Bogo6}
\end{align}
which follow from \eqref{Bogo1}, \eqref{Bogo2} and \eqref{Bogo3}, \eqref{Bogo4}. Using these relations, $a^\dag_n, a_n$ are converted into $b^\dag_n, b_n$ so that they can act on $\ket{0}_\mathrm{M}$. Carrying out the calculation yields
\begin{align}
N_{m,n}=\frac{1}{4\sqrt{\lambda_m\lambda_n}}\sum_{j_1,j_2=1}^N P_{m,j_1}P_{n,j_2}\left[\sqrt{\left(j_1-\frac{1}{2}\right)\left(j_2-\frac{1}{2}\right)}B^{1/2}_{j_1,j_2}-\lambda_n\sqrt{\frac{j_1-\frac{1}{2}}{j_2-\frac{1}{2}}}\delta_{j_1,j_2}\right. \notag\\
\left.-\lambda_m\sqrt{\frac{j_2-\frac{1}{2}}{j_1-\frac{1}{2}}}\delta_{j_1,j_2}+\frac{\lambda_m\lambda_n}{\sqrt{(j_1-\frac{1}{2})(j_2-\frac{1}{2})}}B^{-1/2}_{j_1,j_2}\right].
\end{align}
Numerical evaluation of this expression gives the results shown in Figs.\,\ref{fig3} and \ref{fig4}.

The quantity $N_{n,n}$ deviates significantly from the thermal distribution, and as an example of the case $m\neq n$, $N_{n,n+1}$ takes a non-zero value. As $n$ becomes larger, the magnitude of $N_{n,n+1}$ becomes relatively smaller, which indicates that entanglement between different energy levels in $\ket{0}_\mathrm{M}$ occurs predominantly among low-energy modes. The large deviation of $N_{n,n}$ from the thermal distribution is inevitable once a cutoff is introduced. In the continuum theory, the quantity corresponding to $N_{n,n}$ is given by multiplying the number density $\frac{1}{e^{2\pi\omega/\kappa}-1}$ by $\int_0^\infty\dd{k}\frac{1}{k}\times\mathrm{const}$. This integral diverges not only in the IR region but also in the UV region, and the contribution from sufficiently large $k$ dominates. Therefore, when the calculation is performed with a UV cutoff, the contribution from the region that dominated in the continuum theory is absent, leading to results that differ significantly from those of the continuum theory.

\vskip\baselineskip
However, this result for $N_{n,n}$ does not mean that the thermal property is completely lost when a cutoff is introduced. In a theory with a UV cutoff, the thermal nature disappears only when one attempts to relate the particle number to the particle number density; it does not imply the complete absence of thermal behavior. Therefore, in order to examine a quantity that reflects the thermal property of the Minkowski vacuum while avoiding UV divergences, let us focus on the transition probability $P$ of an Unruh--DeWitt detector. Using the detector-dependent factor $C$ and the energy change measured by the detector $\Delta E$, $P$ is given by
\begin{align}
P=C\int_{-\infty}^\infty \dd{\tau}\int_{-\infty}^\infty\dd{\tau^\prime} e^{-i\Delta E(\tau-\tau^\prime)}{}_\mathrm{M}\mel{0}{\phi(\tau,x)\phi(\tau^\prime,x)}{0}_\mathrm{M}, 
\end{align}
where $\tau,\tau^\prime$ are the proper times of the detector. If the detector has acceleration $\alpha$, one obtains
\begin{align}
\int_{-\infty}^\infty \dd{\tau}\int_{-\infty}^\infty\dd{\tau^\prime} e^{-i\Delta E(\tau-\tau^\prime)}{}_\mathrm{M}\mel{0}{\phi(\tau,x)\phi(\tau^\prime,x)}{0}_\mathrm{M}\propto \frac{\Delta E}{e^{2\pi\Delta E/\alpha}-1}, 
\end{align}
which yields the thermal distribution. The appearance of the thermal factor in this calculation originates from the poles of the Wightman function
\begin{align}
{}_\mathrm{M}\mel{0}{\phi(\tau,x)\phi(\tau^\prime,x)}{0}_\mathrm{M}=-\frac{1}{2\pi}\ln\left[\frac{2\mu}{\alpha}\sinh\left(\frac{\alpha}{2}(\tau-\tau^\prime)\right)\right], 
\label{Wightman}
\end{align}
located at $\tau-\tau^\prime=2\pi in/\alpha\,\,\,(n\in\mathbb{Z})$, from which the Boltzmann factor $e^{-2\pi\Delta E/\alpha}$ arises. ($\mu$ is an IR cutoff.) In other words, the information that the Minkowski vacuum $\ket{0}_\mathrm{M}$ is thermal is encoded in the Wightman function. Therefore, by computing the Wightman function in the lattice theory and comparing it with the continuum result \eqref{Wightman}, we can check whether the Minkowski vacuum retains thermal properties.

\begin{figure}[h]
\centering
\includegraphics[keepaspectratio, scale=0.4]{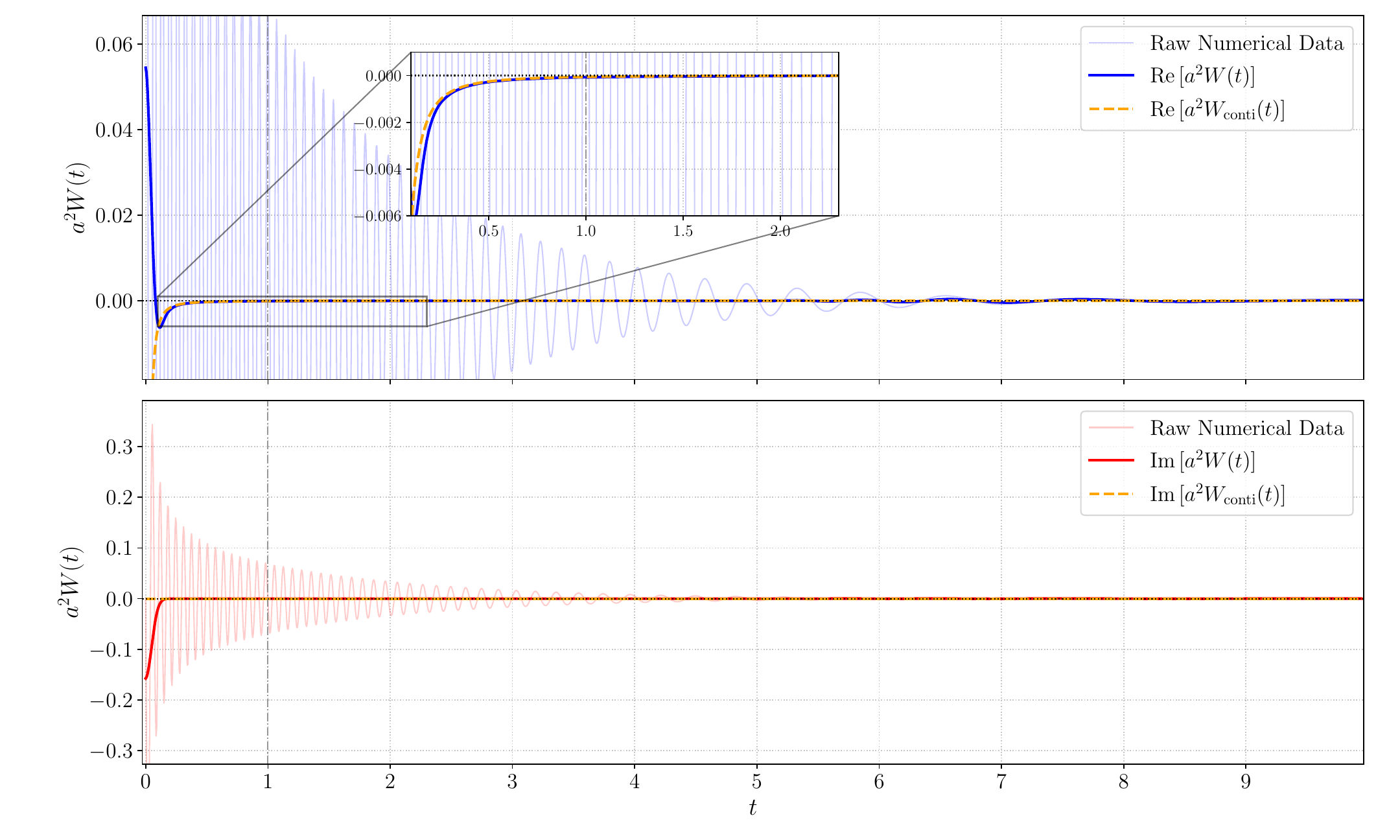}
\caption{Comparison of the Wightman function for $\pi$ between the lattice and continuum theories. We set $N=10000, j_1=50$, and $\kappa=1$. Since the oscillation frequency changes with time $t$, we apply Gaussian smearing with a width of 0.05 for $0<t<1$ and a width of 0.5 for $t\geq 1$.}
\label{fig5}
\end{figure}

\vskip\baselineskip
However, the Wightman function for $\phi$, ${}_\mathrm{M}\mel{0}{\phi(\tau,x)\phi(\tau^\prime,x)}{0}_\mathrm{M}$, again receives a large contribution from the zero mode of $\phi$, which makes numerical calculations unstable. As in \ref{2.3}, we instead compare the Wightman function for $\pi$, ${}_\mathrm{M}\mel{0}{\pi(\tau,x)\pi(\tau^\prime,x)}{0}_\mathrm{M}$. Since the dependence on $\tau-\tau^\prime$ is essential, we may set $\tau^\prime=0$. Fixing $x=x_{j_1}=a(j_1-1/2)$ and using \eqref{Bogo1}, \eqref{Bogo2}, \eqref{Bogo5}, and \eqref{Bogo6} as before, we obtain
\begin{align}
{}_\mathrm{M}\mel{0}{\pi(\tau,x)\pi(0,x)}{0}_\mathrm{M}&=\frac{1}{2a^2}\frac{1}{\sqrt{j_1-\frac{1}{2}}}\sum_{m,j_2=1}^N (P^T)_{j_1,m} P_{m,j_2} \nonumber \\
&\qquad \times\left[\sqrt{j_2-\frac{1}{2}}B^{1/2}_{j_1,j_2}\cos\kappa\lambda_m t-\frac{i}{\sqrt{j_2-\frac{1}{2}}}\delta_{j_1,j_2}\lambda_m\sin\kappa\lambda_m t\right]. \\
&\equiv W(t)
\end{align}
This expression can be evaluated numerically, but in order to compare it with the continuum theory, the resulting waveform needs to be slightly processed. Since we have introduced a UV cutoff, high-energy modes, namely, rapidly oscillating modes, are not represented properly. In other words, in a low-energy effective theory the time resolution is reduced, so in order to extract reliable information from the result we must perform smearing around each value of $t$. After smearing, the resulting Wightman function becomes a quantity that can be compared with the continuum theory. The Wightman function for $\pi$ in the continuum theory, $W_\mathrm{conti}(t)$, is obtained by differentiating \eqref{Wightman} with respect to $t=\frac{\alpha\tau}{\kappa}$ and $t^\prime=\frac{\alpha\tau^\prime}{\kappa}$ and then setting $t^\prime=0$. Using $\alpha=\frac{1}{x}$, we obtain
\begin{align}
W_\mathrm{conti}=-\frac{1}{8\pi}\frac{1}{x^2}\frac{1}{\sinh^2\frac{\kappa t}{2}}, 
\end{align}
and in particular at $x=x_{j_1}=a(j_1-\frac{1}{2})$ this becomes
\begin{align}
W_\mathrm{conti}=-\frac{1}{8\pi}\frac{1}{a^2(j_1-\frac{1}{2})^2}\frac{1}{\sinh^2\frac{\kappa t}{2}}. 
\label{conti}
\end{align}
Normalizing $\kappa=1$ and performing the numerical calculation, we compare the lattice result \eqref{Wightman} with \eqref{conti}, obtaining Fig.\,\ref{fig5}.

As seen from Fig.\,\ref{fig5}, the lattice Wightman function reproduces the continuum result for $t\gtrsim0.5$. On the other hand, there is a large deviation near $t=0$, but this is not problematic because in the continuum theory local operators contain arbitrarily high-energy modes and the Wightman function is not well defined at $t=0$. The discrepancy simply reflects the difference in the UV region between the smeared theory and the unsmeared continuum theory.

\begin{figure}[h]
\centering
\includegraphics[keepaspectratio, scale=0.4]{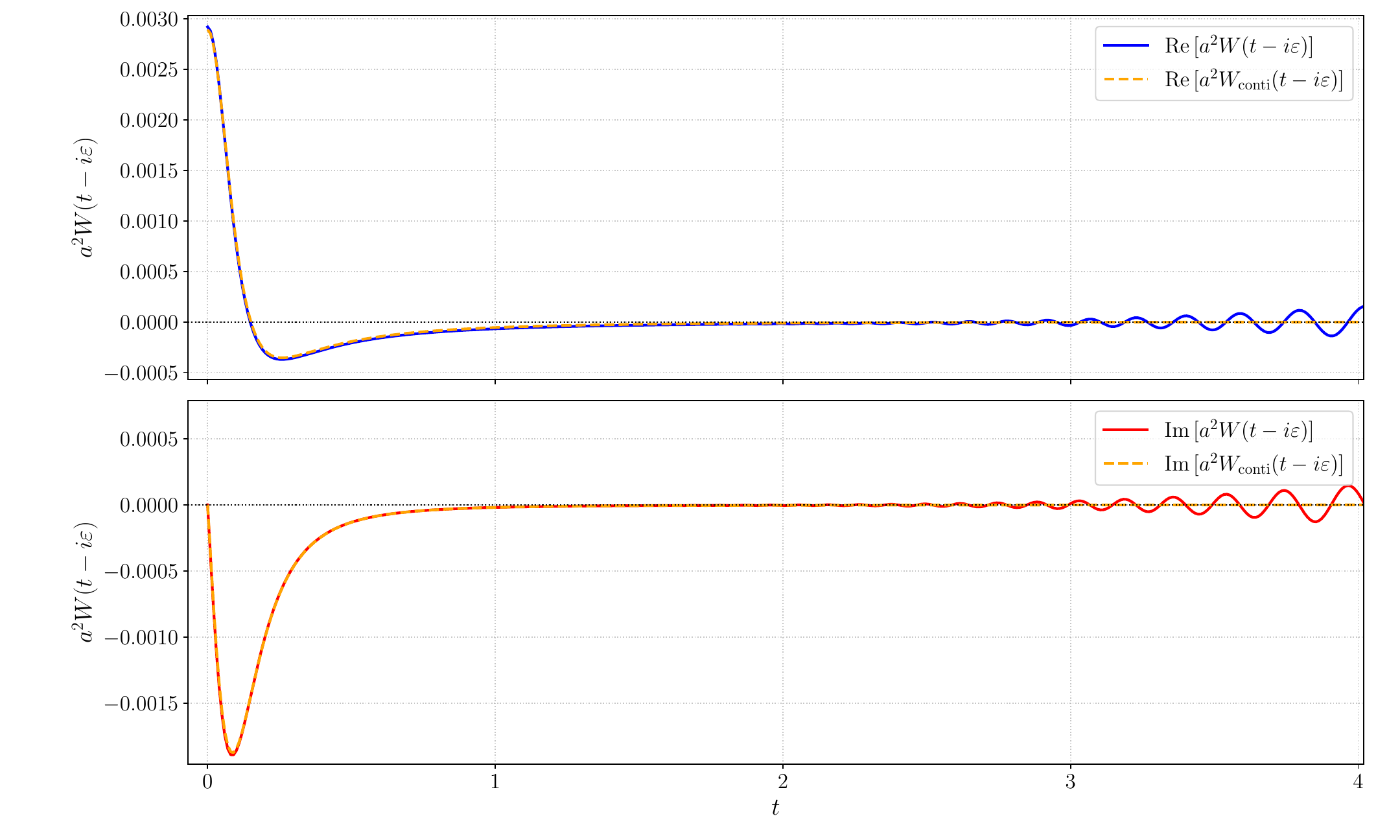}
\caption{Plot of $a^2W(t-i\varepsilon)$ for $N=10000, j_1=50, \kappa=1$, and $\varepsilon=0.15$.}
\label{fig6}
\end{figure}

\vskip\baselineskip
In addition to comparing the smeared Wightman functions, it is also meaningful to compare the Wightman function with suppressed high-frequency contributions by replacing $t\to t-i\varepsilon$ ($\varepsilon$ is a small real number). The result is shown in Fig.\,\ref{fig6}, which demonstrates that the lattice result exhibits the same behavior as $W_\mathrm{conti}(t-i\varepsilon)$ of the continuum theory even near $t=0$. In the region $t\gg1$, low-frequency modes remain, but these agree with the continuum theory once smearing is applied.

\begin{figure}[h]
\centering
\includegraphics[keepaspectratio, scale=0.4]{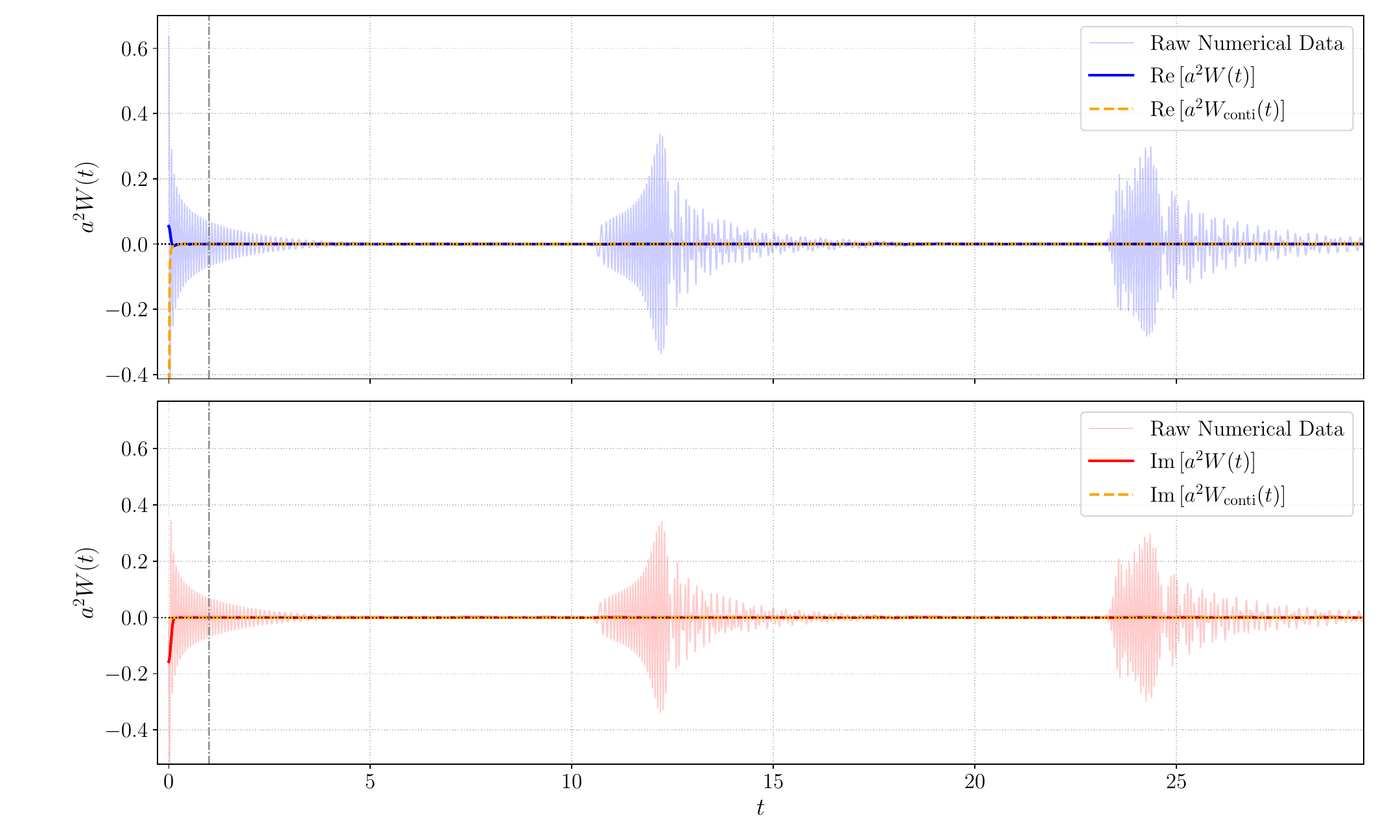}
\caption{A reduced version of Fig.~\ref{fig5}.}
\label{fig7}
\end{figure}

The effect of introducing a UV cutoff in the Wightman function, namely the appearance of the 
brick wall,
%stretched horizon, 
becomes visible for $t \gg 1$. To observe this effect, let us consider Fig.~\ref{fig7}, which is a plot of Fig.~\ref{fig5} with the horizontal axis extended to $t \sim 30$.

To summarize, our numerical results confirm that both the Wightman function and the Unruh-DeWitt detector reproduce the expected thermal behavior, provided that the IR cutoff is sufficiently large, the observation point is far enough from the horizon to avoid causal influence from the brick wall, and the temporal smearing suppresses UV-sensitive high-energy modes.

As in the interpretation of the Green function discussed in \ref{2.3}, the response appearing around $t \sim 11$ corresponds to an outward-propagating wave that is reflected at the boundary $x=x_N$ and returns to $x=x_{j_1}$. The response around $t \sim 12.5$ corresponds to an inward-propagating wave that is reflected at the stretched horizon and returns to $x=x_{j_1}$. The responses appearing around $t \sim 23$ arise from further reflections of these waves at the stretched horizon and the boundary. These features indicate the presence of the stretched horizon.

\subsubsection{Different choice of Rindler Hamiltonian}

Here, we consider the possibility of choosing a Rindler Hamiltonian different from ours. To begin with, the region around 
$n=1$ is highly sensitive to the UV cut-off, so there is certainly room to add field-dependent terms to the Hamiltonian in that region. Even if one does so, however, the qualitative conclusions of this paper should remain unchanged. Such terms do not affect the region far from the horizon, nor can they remove the peculiar behavior induced by the UV cut-off at the brick wall. In this sense, our result is consistent with the intuitive expectation.

By contrast, if one takes the modular Hamiltonian as the Rindler Hamiltonian, then even in a lattice-regularized theory, the Minkowski vacuum becomes a thermofield double state in Rindler coordinates. Here, the state used to construct the modular Hamiltonian is the reduced density matrix obtained by tracing out the left Rindler wedge from the Minkowski vacuum. In this case, however, once the UV cut-off is introduced, the resulting theory is no longer a free theory. Moreover, the Hamiltonian will generally contain nonlocal interactions, in particular couplings between the 
$n=1$ brick-wall region and degrees of freedom far away from it. \footnote{
If one discretizes the local BW ansatz \cite{Bisognano:1976za} \(K_{\mathrm{BW}}^{\mathrm{loc}}\propto \int_{x>0} x\,T_{00}\,\dd{x}\) using the same midpoint prescription as in our paper, then one essentially reproduces our lattice Rindler Hamiltonian, with the kinetic term weighted by \(x_n=(n-\tfrac12)a\) and the gradient term by the link midpoint \(x_{n+\frac12}=na\). Thus, the difference is not in the bulk form, but in the horizon-region ambiguity and, more importantly, in the fact that the exact modular Hamiltonian at finite cutoff is generally nonlocal and is not exhausted by this local BW discretization.
}
Therefore, it is not an appropriate Hamiltonian for describing local Rindler dynamics within a single wedge.

%\section{Discussions}

\section*{Acknowledgement}

The authors would like to thank T. Kawamoto and S. Sugishita for their useful comments.
This work was supported by MEXT-JSPS Grant-in-Aid for Transformative Research Areas (A) ``Extreme Universe'', No. 21H05184.
This work was supported by JSPS KAKENHI Grant Number 	24K07048.

\hspace{1cm}

%Note added:
%As this paper was being completed, we became aware of the preprint \cite{Kinoshita:2023hgc} in which

%\appendix
%\section{XXX}
%\label{a1}

%%%%%%%%%%%%%%%%%%%%%%%%%%%%%%%%%%%%%%%%%%%%%%%%%%%%%%%%%%%%%%
%%%%%%%%%%%%%%%%%%%%%%%%%%%%%%%%%%%%%%%%%%%%%%%%%%%%%%%%%%%%%%
\bibliographystyle{utphys}
\bibliography{main202409.bib}
%%%%%%%%%%%%%%%%%%%%%%%%%%%%%%%%%%%%%%%%%%%%%%%%%%%%%%%%%%%%%%
\end{document}